\documentclass[aps,prb,showpacs,reprint,superscriptaddress]{revtex4-1}
\usepackage[utf8]{inputenc}
\usepackage[pdftex]{graphicx}
\usepackage{amssymb}
\usepackage{physics}
\usepackage{amsmath}
\usepackage{xfrac}
\usepackage{graphicx}
\usepackage{dcolumn}
\usepackage{color}
\usepackage{bm}
\usepackage{hyperref}

\newcommand{\ems}{EuMnSb$_2$}
\begin{document}
\preprint{APS/123-QED}
	
\title{Magnetic and electronic structure of Dirac semimetal candidate \ems}
\author{J.-R. Soh}
\affiliation{Department of Physics, University of Oxford, Clarendon Laboratory, Parks Road, Oxford OX1 3PU, UK}
\author{P. Manuel}
\affiliation{ISIS Pulsed Neutron Facility, STFC, Rutherford Appleton Laboratory, Chilton, Didcot, Oxfordshire OX11-0QX, UK}
\author{N. M. B. Schr\"oter}
\affiliation{Swiss Light Source, Paul Scherrer Institute, CH-523 2 Villigen PSI, Switzerland}
\author{C. J. Yi}
\affiliation{Beijing National Laboratory for Condensed Matter Physics, Institute of Physics, Chinese Academy of Sciences, Beijing 100190, China}
\author{D. Prabhakaran}
\affiliation{Department of Physics, University of Oxford, Clarendon Laboratory, Parks Road, Oxford OX1 3PU, UK}
\author{F. Orlandi}
\affiliation{ISIS Pulsed Neutron Facility, STFC, Rutherford Appleton Laboratory, Chilton, Didcot, Oxfordshire OX11-0QX, UK}
\author{Y. G. Shi}
\affiliation{Beijing National Laboratory for Condensed Matter Physics, Institute of Physics, Chinese Academy of Sciences, Beijing 100190, China}
\author{A. T. Boothroyd}
\email{andrew.boothroyd@physics.ox.ac.uk}
\affiliation{Department of Physics, University of Oxford, Clarendon Laboratory, Parks Road, Oxford OX1 3PU, UK}%
\date{\today}
	
\begin{abstract}
We report an experimental study of the magnetic order and electronic structure and transport of the layered pnictide \ems, performed using neutron diffraction,  angle-resolved photoemission spectroscopy (ARPES), and magnetotransport measurements. We find that the Eu and Mn sublattices display antiferromagnetic (AFM) order below  $T_\mathrm{N}^\mathrm{Eu} = 21(1)$\,K and $T_\mathrm{N}^\mathrm{Mn} = 350(2)$\,K respectively. The former can be described by an A-type AFM structure with the Eu spins aligned along the $c$ axis (an in-plane direction), whereas the latter has a C-type AFM structure with Mn moments along the $a$--axis (perpendicular to the layers). The ARPES spectra reveal Dirac-like linearly dispersing bands near the Fermi energy. Furthermore, our magnetotransport measurements show strongly anisotropic magnetoresistance, and indicate that the Eu sublattice is intimately coupled to conduction electron states near the Dirac point. 
	\end{abstract}
	
	
	\maketitle
	Topological semimetals can host quasiparticle excitations which masquerade as massless fermions due to the linearly-dispersing electronic bands created by interactions with the crystal lattice. The Dirac or Weyl nodes, where the conduction and valence bands meet in momentum space, are robust against small perturbations due to the protection afforded by crystalline symmetries or the topology of the electronic bands~\cite{Armitage2018,Burkov2016,Rau2016,Hasan2010,Pesin2009}. Topological semimetals exhibit exceptional electronic transport properties (e.g. extremely high carrier mobility and large linear magnetoresistance) and the control of these exotic charge carriers could help realize a new generation of spintronic devices with low power consumption~\cite{mejkal2017,Smejkal2017,Shi2015}. 
	
\begin{figure}[b]
		\includegraphics[width=0.5\textwidth]{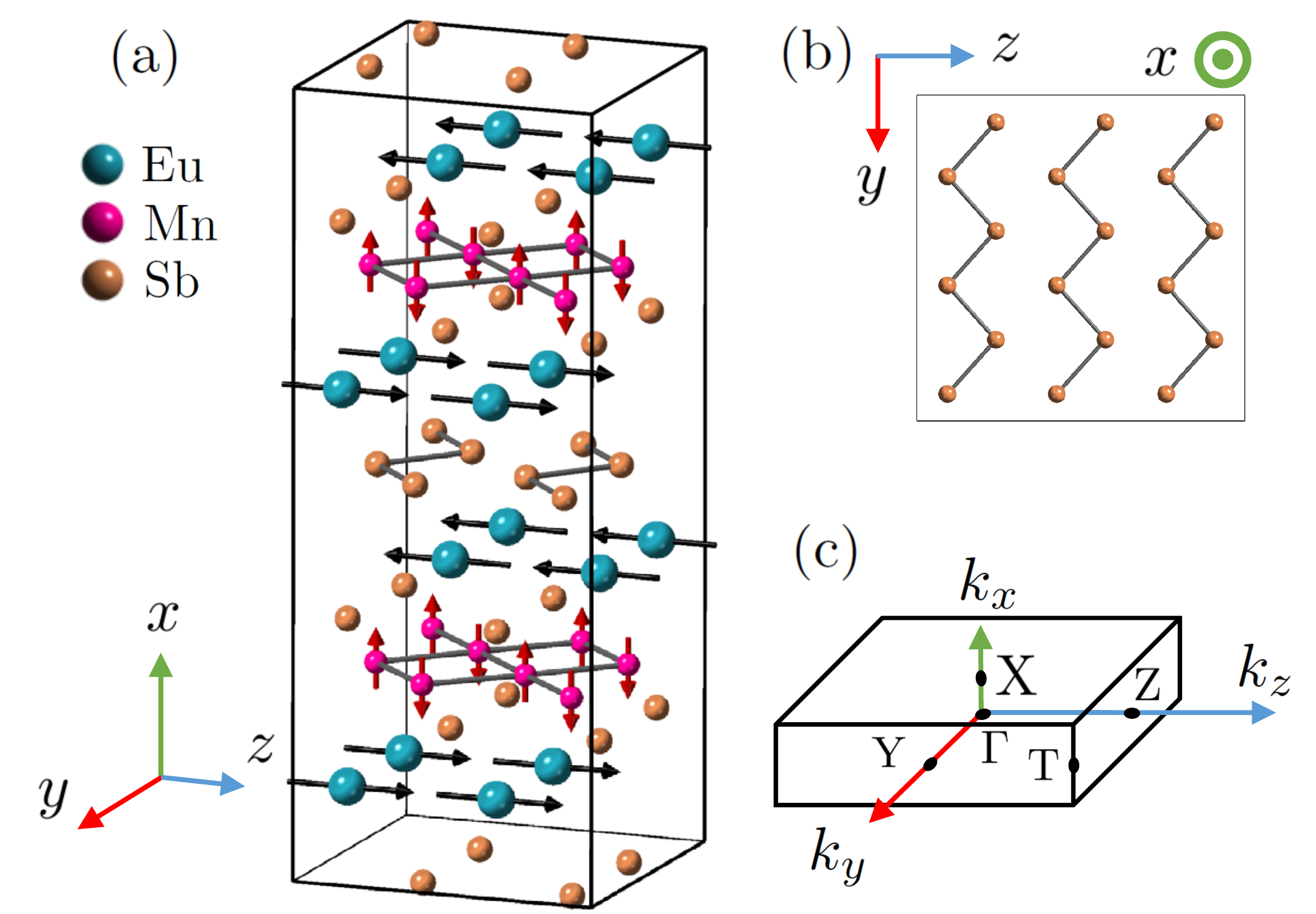}
		\caption{ (a) The crystal and magnetic structure of \ems\, ($1\times2\times2$ unit cells). The Eu and Mn sublattices exhibit A-type and C-type AFM order, respectively. (b) View down the $x$ axis of the orthorhombic unit cell (space group $Pnma$, lattice parameters $a$ =  22.567(5)\,\AA, $b$= 4.371(1)\,\AA, $c$ =  4.409(1)\,\AA\, at $T=300$\,K) showing the zig-zag network of Sb atoms in the $y$--$z$ plane. (c) High symmetry points of the orthorhombic Brillouin zone. 
		\label{fig:CS12a_EuMnSb2_Images}}
	\end{figure}

	Such control can potentially be realized in materials in which magnetic order coexists with non-trivial electronic band topology. Recent ARPES, quantum oscillation, neutron diffraction and \textit{ab initio} band structure studies suggest that materials in the \textit{A}MnSb$_2$ (\textit{A} = Ca, Sr, Ba, Eu, Yb) family display many of the required properties~\cite{Liu2016,Farhan2014,Kealhofer2018,Wang2018,Ramankutty2018,Liu2017b,He2017,You2019,Yan2017,Soumya2017}. The two-dimensional zig-zag layer of Sb atoms [Fig.~\ref{fig:CS12a_EuMnSb2_Images}] in these 112--pnictides play host to fermions which can be described by the relativistic Dirac or Weyl equations. Furthermore, the electronic transport in this family of materials also displays large magnetoresistive effects, suggesting a coupling between the magnetism and charge carriers~\cite{Liu2016,Farhan2014,Kealhofer2018,Wang2018,Ramankutty2018,Liu2017b,He2017,You2019,Yan2017}. These effects could be driven by changes in the electronic band structure topology due to changes in the symmetry of the spin structures induced by the applied field~\cite{Klemenz2019}.

	Within the $A$MnSb$_2$ family, \ems\, is of particular interest because the conducting zig-zag layer of Sb atoms is sandwiched between two interpenetrating magnetic sublattices (Eu and Mn), as shown in Fig.~\ref{fig:CS12a_EuMnSb2_Images}(a). Such a structure may lead to an enhancement of the coupling between the topological quasiparticles and magnetism, compared to that in compounds with a non-magnetic atom on the $A$ site.  The dramatic magnetoresistive behaviour observed in a recent work~\cite{Yi2017} is evidence for the importance of this coupling. Up to now, however, the nature of the magnetic order in EuMnSb$_2$, which determines  the character of any topological band crossings, has yet to be determined. Moreover, although the magnetic susceptibility of \ems\, has an anomaly at 20\,K, which is attributed to the AFM order of Eu, no evidence for magnetic order of the Mn spins has been detected up to 400\,K.~\cite{Yi2017} This is surprising given that all other \textit{A}MnX$_2$ ($X = {\rm Sb, Bi})$ systems studied thus far exhibit AFM order of Mn near room temperature.~\cite{Wang2012,Wang2016,Liu2016,Farhan2014,Kealhofer2018,Wang2018,Ramankutty2018,Liu2017b,He2017,Brechtel1981,Park2011,Feng2014,Masuda2016,Jo2014,Borisenko2015,Liu2017,May2014,Lee2013,Guo2014,Chinotti2016,Wang2011,You2019,Yan2017,He2011}

	In light of this, we set out in this study to (i) determine the spin configuration of the two magnetic sublattices by powder neutron diffraction, (ii) investigate the nature of electronic states by mapping the band structure with ARPES, and (iii) shed light on the coupling between magnetism and the charge carriers through magnetotransport measurements. We find that the Mn sublattice displays long range order below $T_\mathrm{N}^\mathrm{Mn} = 350(2)$\,K with the magnetic moments aligned along the $a$ axis in a C-type AFM structure. Upon further cooling, we find that the Eu spins exhibit an A-type AFM ordering at $T_\mathrm{N}^\mathrm{Eu} = 21(1)$\,K with the spins aligned along the $c$ axis. Our ARPES results show that \ems\, displays Dirac-like linearly dispersing electronic bands near the Fermi energy. Furthermore, we demonstrate that these electronic states are strongly coupled to the Eu magnetic sublattice, evidenced by the magnetotransport data presented in this work.
	
	Bulk \ems\, single crystals were grown via the flux method, as described in Ref.~\onlinecite{Yi2017}. The unit cell can be described by the $Pnma$ space group with the Eu and Mn atoms at two symmetry-inequivalent 4$c$ Wyckoff  positions $(0.38637(2),0.25,0.76977(5))$ and $(0.24970(3),0.25,0.27005(14))$, respectively ($T = 300$\,K). Magnetotransport measurements were performed on a Quantum Design Physical Properties Measurement System (PPMS) for fields up to 13\,T in a Hall configuration with the standard five-contact method. The longitudinal component, $\rho_{zz}$, was obtained by symmetrizing the data measured in positive and negative fields respectively. Here we define $\rho_{zz} = E_z/j_z $, where $E_z$ and $j_z$ are the electric field and current density along $z$ respectively.
	
	 Powder neutron diffraction measurements of \ems\, were performed on the high-flux diffractometer WISH (ISIS Facility, UK)~\cite{Chapon2011}. The \ems\, single crystals were crushed in argon and loaded in a vanadium can. The can had a relatively small diameter (3\,mm) to reduce the attenuation due to the high neutron absorption of Eu ($\sigma_{\rm a}^{\mathrm{Eu}}$ = 4530 barns). 
	 ARPES spectra of \ems\, were recorded with soft x-rays on the SX-ARPES end station of the ADRESS beamline (Swiss Light Source, Switzerland)~\cite{Strocov2010}. Measurements were performed with a SPECS analyzer at a photon energy of 790 eV with right-circularly polarized light $(C+)$. The sample was cleaved normal to the [100] direction and measured at $T\sim$ 20\,K at a vacuum better than $1\times10^{-10}$ mbar. 

	\begin{figure}[t]
		\includegraphics[width=0.5\textwidth]{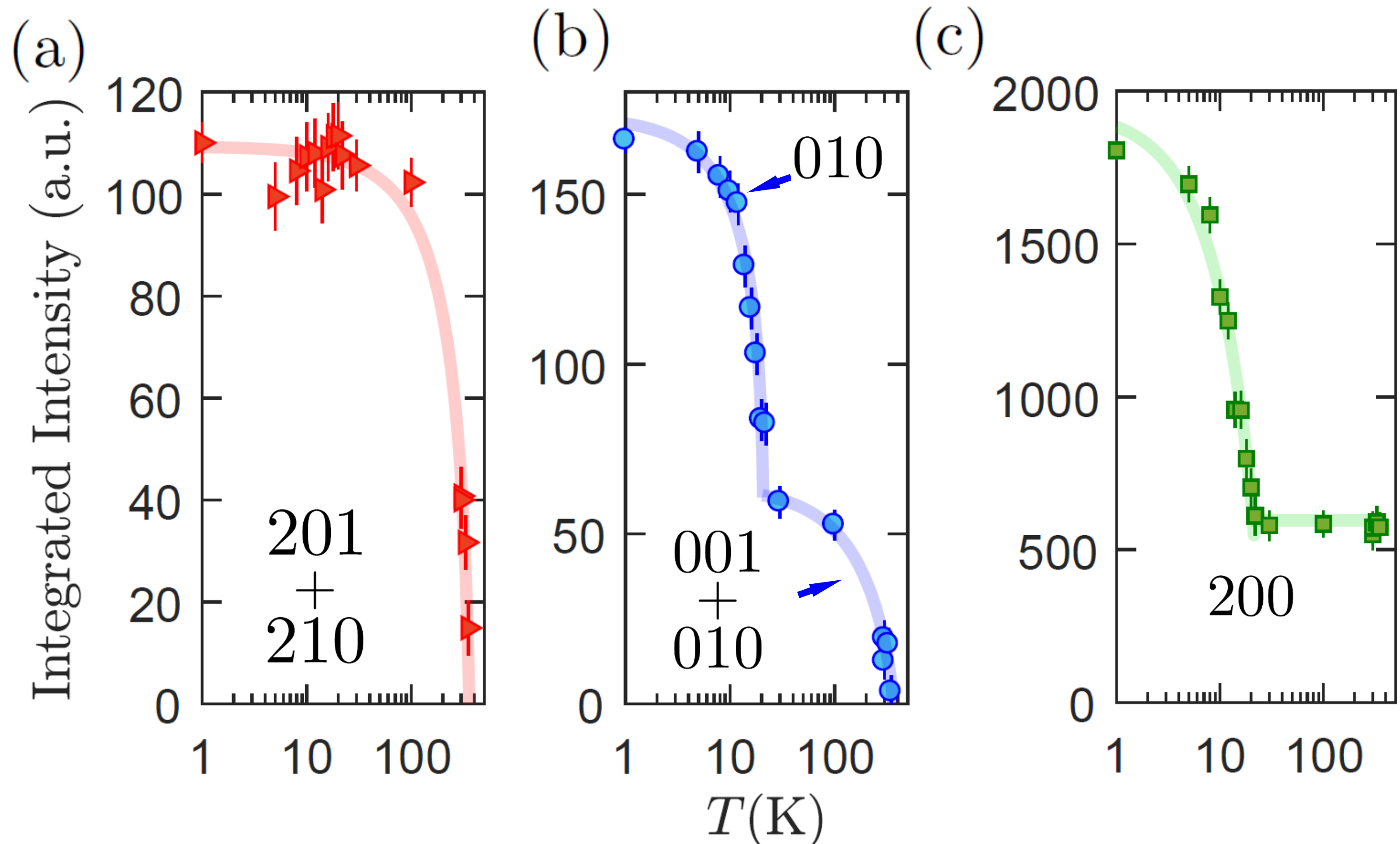}
		\includegraphics[width=0.45\textwidth]{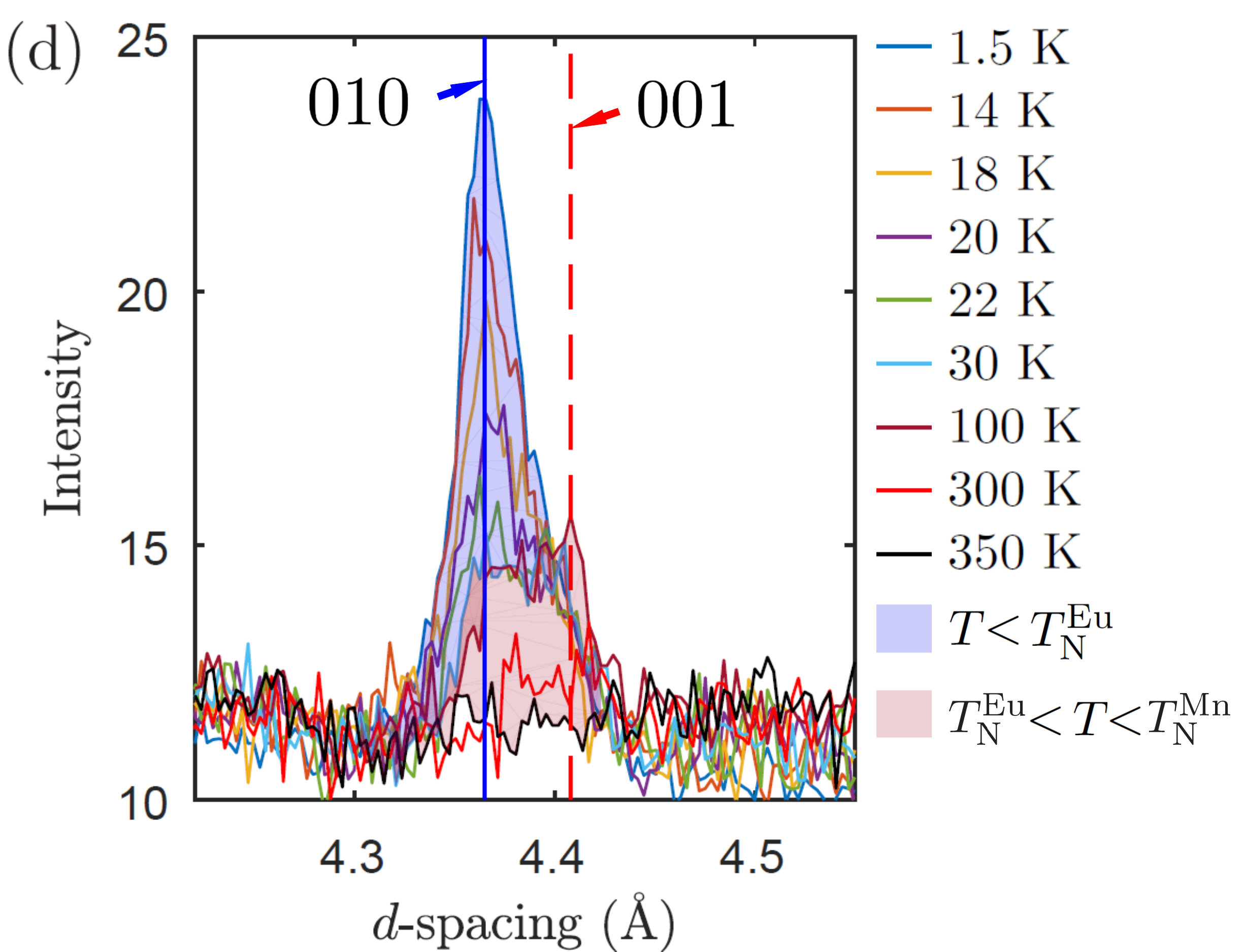}
		\caption{\label{fig:CS12b_EuMnSb2_WISH} (a)--(c) Temperature dependence (log scale) of the integrated intensity of the peaks at $d$-spacings 4.09\,\AA, 4.38\,\AA\, and 11.25\,\AA\,, respectively. (d) For $T_\mathrm{N}^\mathrm{Eu}<T<T_\mathrm{N}^\mathrm{Mn}$, the purely magnetic peak at 4.38\AA\, comprises the peaks $010$ and $001$. Below $T_\mathrm{N}^\mathrm{Eu}$, only the intensity of the $010$ reflection grows.}
	\end{figure}

Figure~\ref{fig:CS12b_EuMnSb2_WISH} shows the temperature dependence of three peaks, with $d$-spacings of 4.09\,\AA, 4.38 \,\AA\, and 11.25\,\AA\,, respectively. The intensity of the first two peaks increases sharply below $T_\mathrm{N}^\mathrm{Mn} = 350(2)$\,K, signalling the onset of magnetic order on the Mn sublattice. The intensity at $4.09$\AA\, continues to grow and saturates as the temperature approaches 1.5\,K. On the other hand, the intensity of the reflection at $4.38$\AA\, displays another sharp increase below $T_\mathrm{N}^\mathrm{Eu} = 21$\,K, which is also observed at 11.25\,\AA. This transition is associated with the magnetic ordering of the Eu moments, and coincides with the anomaly in the temperature-dependent magnetotransport data (see later).

For $T_\mathrm{N}^\mathrm{Eu}<T<T_\mathrm{N}^\mathrm{Mn}$, the peak at 4.38\,\AA\, comprises two reflections at slightly different $d$--spacing, namely $4.37$\AA\, and $4.41$\AA, which can be indexed by the Miller indices $010$ and $001$. These peaks are not well separated as the ratio $c/b=4.41/4.37$ is close to unity [See Fig.~\ref{fig:CS12b_EuMnSb2_WISH}(d)]. The $010$ and $001$ peaks are purely magnetic as they are forbidden nuclear reflections in the $Pnma$ space group. The peak at 4.09\,\AA\, is similarly made up of 2 reflections, $201$  and $210$, at $d$-spacings 4.10\AA\, and 4.08\AA\,, respectively. These  are weak nuclear reflections in the paramagnetic phase which display additional magnetic intensity below $T_\mathrm{N}^\mathrm{Mn}$.   
    
The peak at 11.25\,\AA, which is the nuclear-allowed $200$ reflection, displays additional  intensity due to magnetic scattering only for $T<T_\mathrm{N}^\mathrm{Eu}$, Fig.~\ref{fig:CS12b_EuMnSb2_WISH}(c).
The two magnetic orderings are also represented in Fig.~\ref{fig:CS12b_EuMnSb2_WISH}(d), which shows the temperature dependence of the composite magnetic peak at 4.38\,\AA.  As the intensity of the $001$ component at 4.41\,\AA\, is already saturated by 100\,K, the increase in the integrated intensity for 
$T<T_\mathrm{N}^\mathrm{Eu}$ [Fig.~\ref{fig:CS12b_EuMnSb2_WISH}(b)] only arises from the $010$ peak at 4.37\,\AA.  Additional peaks, which can be indexed by $211$, $401$, $410$, $600$ and $400$, were also observed below $T_\mathrm{N}^\mathrm{Eu}$.

All of the observed magnetic peaks are consistent with the magnetic propagation vector of $\textbf{k}=\textbf{0}$ for both Mn and Eu. As both magnetic species have the same magnetic propagation vector and are on the same Wyckoff site (but different positions in the lattice), the magnetic order on both sublattices can be described by the same set of irreps. The magnetic $\Gamma$--point representation for the 4$c$ Wyckoff position $(x,0.25,z)$ of the $Pnma$ space group decomposes into eight distinct one-dimensional irreducible representations (irreps), $\Gamma = \Gamma_1^+ + 2\Gamma_1^- + 2\Gamma_2^+ + \Gamma_2^- + 2\Gamma_3^+ + \Gamma_3^- + \Gamma_4^+ + 2\Gamma_4^-$, with the associated basis vectors listed in Table~\ref{tab:table1} (the Miller and Love notation is adopted \cite{Miller1967}). In both cases, the irreps which describe the ferromagnetic sublattices with moments along $x$, $y$ and $z$, namely $\Gamma_3^+$, $\Gamma_4^+$ and $\Gamma_2^+$, can be excluded by the bulk magnetization~\cite{Yi2017}. We consider all of the remaining irreps for the Eu and Mn sublattices systematically and discuss the magnetism on each in turn.  

\begin{table}[b]
\caption{\label{tab:table1}
Representational analysis for the $4c$ Wyckoff site 
of the $Pnma$ space group ($\textbf{k}=\textbf{0}$). The first column shows the relative spin arrangement at the atomic positions $(x,0.25,z)$, $(-x+0.5,0.75,z+0.5)$, $(-x,0.75,-z)$ and $(x+0.5,0.25,-z+0.5)$ respectively. The last column denotes whether the irrep can be described by one of the A$_i$, C$_i$ or G$_i$ types of AFM, or by a ferromagnet, F$_i$, where $i = x,y,z$ indicates the spin orientation.}
\begin{ruledtabular}
\begin{tabular}{lcccr}
Basis&Orien-&Irrep&\multicolumn{2}{c}{Structure}\\
vector&tation  && Eu &Mn\\[2pt]
\colrule
         &$x$&$\Gamma_3^+$ &F$_x$&F$_x$ \\
$(++++)$ &$y$&$\Gamma_4^+$ &F$_y$&F$_y$ \\
         &$z$&$\Gamma_2^+$ &F$_z$&F$_z$\\
\hline
         &$x$&$\Gamma_4^-$ &     &A$_x$ \\
$(++--)$ &$y$&$\Gamma_3^-$ &     &A$_y$\\
         &$z$&$\Gamma_1^-$ &     &A$_z$\\
\hline
         &$x$&$\Gamma_2^+$ &     &G$_x$\\
$(+-+-)$ &$y$&$\Gamma_1^+$ &     &G$_y$\\
         &$z$&$\Gamma_3^+$ &     &G$_z$ \\
\hline
         &$x$&$\Gamma_1^-$ &A$_x$&C$_x$\\
$(+--+)$ &$y$&$\Gamma_2^-$ &A$_y$&C$_y$ \\
         &$z$&$\Gamma_4^-$ &A$_z$&C$_z$ \\
 \end{tabular}
\end{ruledtabular}
\end{table}

For the Mn sublattice, the basis vector $(++--)$  [See Table~\ref{tab:table1}] which describes A-type AFMs ($\Gamma_4^-$, $\Gamma_3^-$ and $\Gamma_1^-$), does not produce the observed $201$, $210$, $001$ and $010$ magnetic reflections. On the other hand, the G-type AFMs [$(+-+-)$ basis vector] with Mn moments along $x$, $y$ and $z$ do not give rise to the $210$ and $010$ magnetic reflections. The basis vector $(+--+)$ with Mn moments along $y$ and $z$, is not predicted to have $010$ and $001$ magnetic reflections, respectively. Hence, the only irrep that is consistent with the observed peaks is the $\Gamma_1^-$ irrep, which can be described by a C--type AFM with Mn moments along $x$. 

For the Eu sublattice, the basis vector $(++--)$ with moments along $x$, $y$ and $z$, corresponding to the irreps $\Gamma_4^-$, $\Gamma_3^-$ and $\Gamma_1^-$ respectively, do not produce the observed $010$ and $200$ magnetic reflections. The same is true for the basis vector $(+-+-)$ with irreps $\Gamma_2^+$, $\Gamma_1^+$ and $\Gamma_3^+$. For the basis vector $(+--+)$, the irreps with Eu moments along $x$ and $z$ predict additional intensity at $001$ which is not observed below $T_\mathrm{N}^\mathrm{Eu}$. This leaves us with an A-type AFM with Eu moments along $y$ which can be described by the $\Gamma_2^-$ irrep. 

For $T_\mathrm{N}^\mathrm{Eu}<T< T_\mathrm{N}^\mathrm{Mn}$, the symmetry of the \ems\, crystal can be described by the $Pn'm'a'$ Shubnikov group due to the Mn order. This symmetry is lowered to $P2_1/a'$ [unique axis $c$, $\gamma \neq 90^\circ$], below the Eu ordering temperature, due to the combined effect of the two irreps. The magnetic structures for the Mn and Eu sublattices are presented in Fig.~\ref{fig:CS12a_EuMnSb2_Images}(a).

	\begin{figure}[t]
		\includegraphics[width=0.5\textwidth]{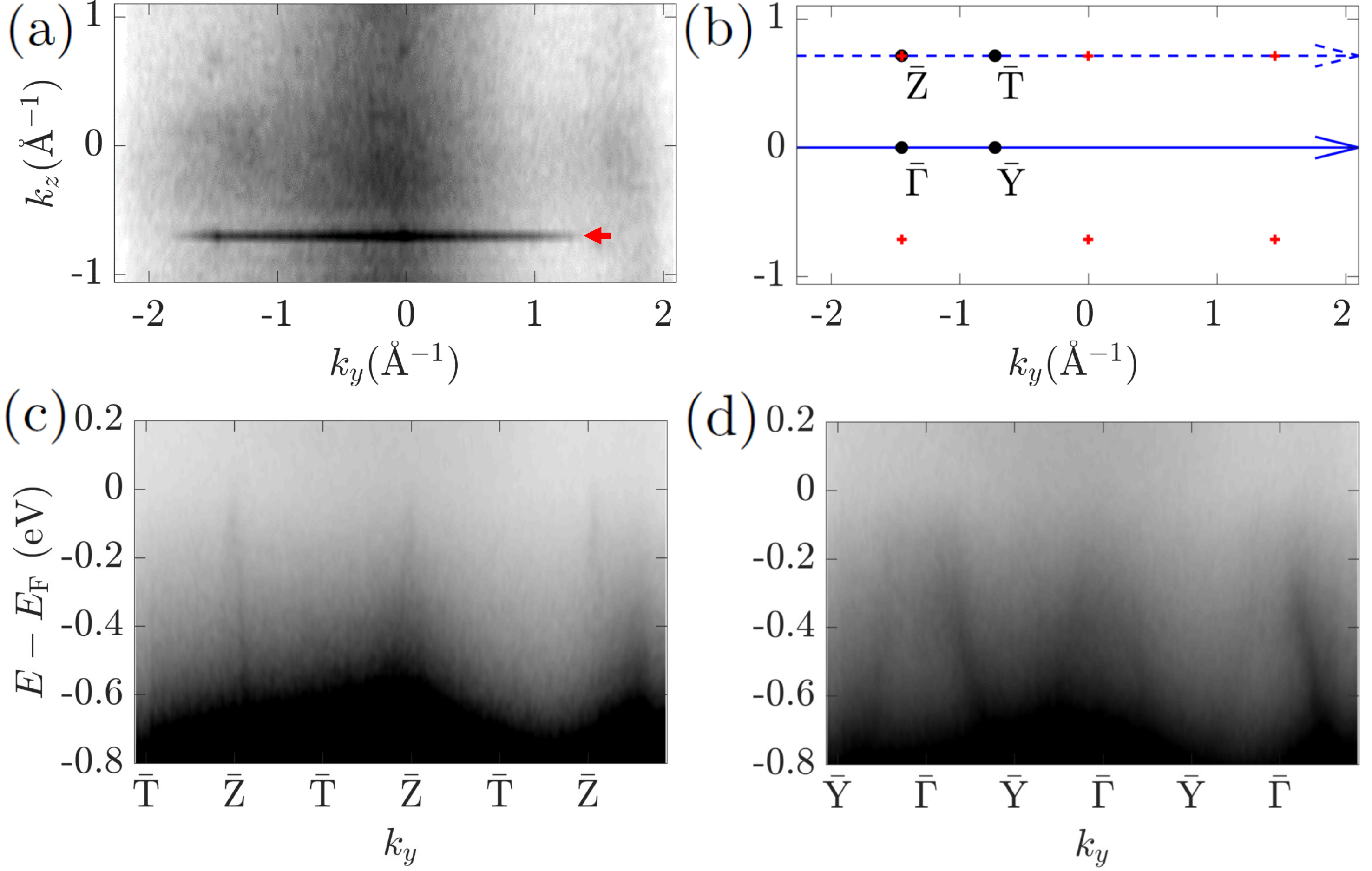}
		\caption{(a) Constant energy surface measured at 790 eV with right--circularly polarized light at $E=E_\mathrm{F}$. (b) High symmetry points in the $k_y$--$k_z$ plane, as defined for the BZ shown in Fig.~\ref{fig:CS12a_EuMnSb2_Images}(c). (c), (d) Electronic dispersion along the high symmetry line cuts indicated in (b) (dashed and solid blue arrows, respectively). Note that the streak in (a), denoted by the red arrow, is a detector artifact. \label{fig:CS12d_EuMnSb2_ARPES}}
	\end{figure}

	Figures~\ref{fig:CS12d_EuMnSb2_ARPES}(a), (c) and (d) illustrate our ARPES measurements of the electronic structure of \ems, which were made at $T\simeq$ 20\,K. The constant energy map in the $k_y$--$k_z$ plane at 0 eV binding energy (where $E = E_\mathrm{F}$), as shown in  Fig.~\ref{fig:CS12d_EuMnSb2_ARPES}(a), reveals small Fermi pockets at the $\bar{\mathrm{Z}}$ high symmetry points across the six Brillouin zones [marked by the red crosses in Fig.~\ref{fig:CS12d_EuMnSb2_ARPES}(b)]. Furthermore, the ${\bf k}$--$E$ dispersion slice in Fig.~\ref{fig:CS12d_EuMnSb2_ARPES}(c) for $\bf k$ along the $\bar{\mathrm{Z}}$--$\bar{\mathrm{T}}$--$\bar{\mathrm{Z}}$ high symmetry line [depicted by the dashed blue arrow in Fig.~\ref{fig:CS12d_EuMnSb2_ARPES}(b)] shows steeply-dispersing linear bands converging at the $\bar{\mathrm{Z}}$ high symmetry points at $E_\mathrm{F}$. This provides direct evidence for Dirac fermions which could be important for electronic transport.
	
	 Figure~\ref{fig:CS12d_EuMnSb2_ARPES}(d), which plots the ${\bf k}$--$E$ dispersion for $\bf k$ along the $\bar{\mathrm{\Gamma}}$--$\bar{\mathrm{Y}}$--$\bar{\mathrm{\Gamma}}$ line cut, as depicted by the solid blue arrow in Fig.~\ref{fig:CS12d_EuMnSb2_ARPES}(b), shows linearly dispersing bands that terminate at $E_\mathrm{F}$. These bands, unlike those in Fig.~\ref{fig:CS12d_EuMnSb2_ARPES}(c), do not converge at $E_\mathrm{F}$, but instead form  hole  pockets  centered around the $\bar{\Gamma}$ point.
	
	
	Furthermore, we also observe a smearing out of the bands beyond binding energies of about 0.5\,eV and 0.7\,eV in Figs.~\ref{fig:CS12d_EuMnSb2_ARPES}(c) and (d), respectively. This effect can arise from the Eu 4$f$ bands which are strongly correlated due to the highly localized nature of the orbitals. Hence, ARPES shows that the $4f$ states, which give rise to the Eu magnetism, are not expected to play a direct r\^{o}le in electronic transport. Our magnetotransport data, which we will now show, demonstrate that the $4f$ electrons can nonetheless influence charge transport indirectly through exchange interactions.
		
\begin{figure}[t]
		\includegraphics[width=0.5\textwidth]{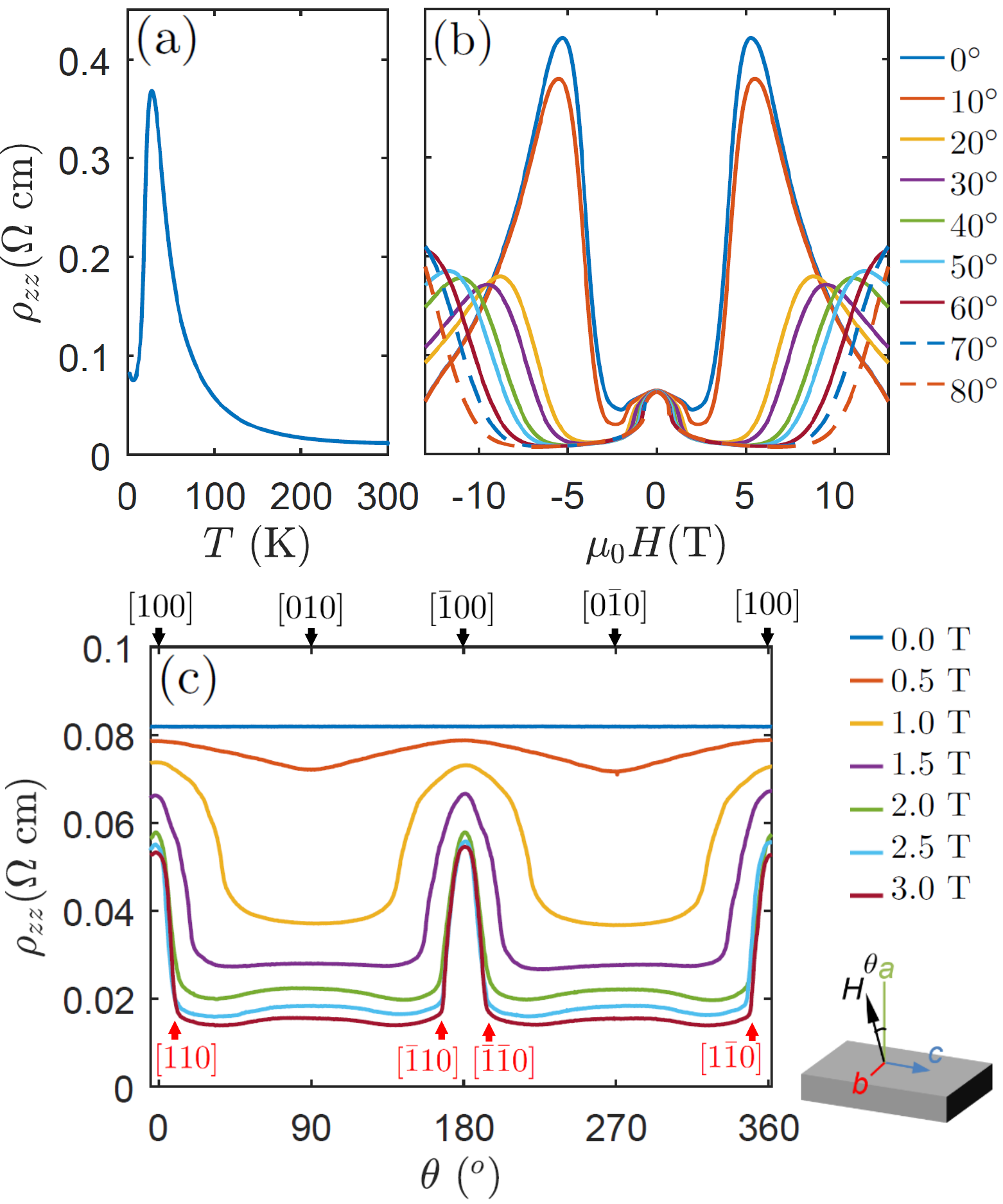}
		\caption{(a) The in--plane resistivity shows a metal-to-insulator transition which peaks at $T_\mathrm{N}^{\mathrm{Eu}} \sim 20$\,K before falling at lower temperatures. (b) Field dependent $\rho_{zz}$ at various field directions measured at $T$ = 2\,K. Here the angles [see insert of Fig.~\ref{fig:CS12c_EuMnSb2_PPMS}(c)] are defined with respect to the $x$ axis where the field direction is fixed to be within the $x$--$y$ plane. (c) Angular dependence of $\rho_{zz}$ at various field strengths demonstrate the strong anisotropy in the magneto-transport of EuMnSb$_2$.
		\label{fig:CS12c_EuMnSb2_PPMS} }
	\end{figure}

	 The longitudinal magnetotransport behavior of \ems\, with the current along $z$ is summarized in Fig.~\ref{fig:CS12c_EuMnSb2_PPMS}. Figure~\ref{fig:CS12c_EuMnSb2_PPMS}(a) plots $\rho_{zz}$ as a function of temperature in zero field. The resistivity increases strongly on cooling, reaching a maximum at $T_\mathrm{N}^{\mathrm{Eu}}$ with $\rho_{zz}^{20\,\mathrm{K}}/\rho_{zz}^{300\,\mathrm{K}}\sim31$, then decreases at lower temperatures. This behavior is consistent with paramagnetic scattering of conduction electrons by the Eu spins, and is evidence  that the charge carriers in \ems\, are strongly coupled to spin fluctuations associated with the magnetic ordering on the Eu sublattice.
	
	To help understand the magnetoresistance behavior further, we also measured $\rho_{zz}$ at $T=2$\,K as a function of applied field in several directions within the $x$--$y$ plane, where the angle $\theta$ is defined with respect to the crystallographic $a$ axis [see insert of Fig.~\ref{fig:CS12c_EuMnSb2_PPMS}(c)]. The data are plotted in Fig.~\ref{fig:CS12c_EuMnSb2_PPMS}(b). At $\theta = 0^\circ$, three distinct features can be identified in $\rho_{zz}$: (i) In the low field regime ($\mu_0H \lesssim$ 3 T), there is a small drop in $\rho_{zz}$. (ii) In the intermediate regime (3 T $\lesssim \mu_0H \lesssim$ 5.3 T) there is a sharp increase in $\rho_{zz}$, with the fractional change in $\rho_{zz}(H)$ reaching 9.4 at the maximum. (iii) In the high field regime ($\mu_0H > 5.3 $\,T), the anomalous resistivity decreases again with increasing field.
	
	With increasing field applied along the $[100]$ direction ($\theta = 0$) the AFM-ordered Mn spins are expected to flop into the $y$--$z$ plane. Subsequently, both the Mn and Eu spins will rotate towards the $x$ axis and eventually become fully spin-polarized at their respective saturation fields. Strong spin fluctuations are expected during this evolution in the magnetic structures, which can lead to enhanced scattering of the conduction electrons. An exchange-induced modification of the electronic bands is also possible,  arising out of the changes in the magnetic symmetry due to the field induced alteration of spin ordering which can lift or add symmetry protection to the band crossing at high symmetry points in the Brillouin zone~\cite{Schoopeaar2317,Vergniory2018}. These are two possible mechanisms which could account for the large magnetoresistive effects seen in Fig.~\ref{fig:CS12c_EuMnSb2_PPMS}(b).

	The $\rho_{zz}$ magnetoresistance  displays strong angular dependence, with the peak moving from $5.3$\,T for $\theta = 0^\circ$, to about $8$\,T for $\theta = 20^\circ$, and eventually beyond the measurement window range of $13$\,T, for $\theta = 80^\circ$. For  $\mu_0H = \pm 5.3$\,T, we find that the ratio of $\rho_{zz}$ measured at $\theta = 0^\circ$ to that at $\theta = 80^\circ$, or $\rho_{zz}^{0^\circ}/\rho_{zz}^{80^\circ}$, is 54. Another view of this anisotropy is given in  Fig.~\ref{fig:CS12c_EuMnSb2_PPMS}(c), which plots $\rho_{zz}$ as a function of $\theta$, measured at $T=2\,$K at various field strengths up to 3\,T. For this measurement the in-plane resistivity was continuously measured as the applied field was rotated about the current $j_z$. The field thus passes through the principal crystal axes $[100]$, $[010]$, $[\bar{1}00]$ and $[0\bar{1}0]$ at $\theta = 0^\circ$, $90^\circ$,  $180^\circ$ and $270^\circ$, respectively [see insert of Fig.~\ref{fig:CS12c_EuMnSb2_PPMS}(c)]. The $180^\circ$ periodicity in $\rho_{zz}$ is in agreement with charge transport in a 2-dimensional layer. Moreover, consistent with the $|\mu_0H| \leq$\,3\,T data presented in Figure~\ref{fig:CS12c_EuMnSb2_PPMS}(b), $\rho_{zz}$ decreases in all field directions, forming sharp resistivity peaks at  angles  $\theta = 0^\circ$, $180^\circ$ and $360^\circ$. At $\mu_0H = 3$\,T, the peaks have an angular width of $\delta\theta\sim20^\circ$, and are suppressed when $\theta=10^\circ$, $170^\circ$, $190^\circ$ and $350^\circ$, as denoted by the red arrows in Fig.~\ref{fig:CS12c_EuMnSb2_PPMS}(c). Intriguingly, these angles correspond to the $[110]$, $[\bar{1}10]$, $[\bar{1}\bar{1}0]$  and $[1\bar{1}0]$ directions of the orthorhombic crystal respectively.
	
	The results  presented in Figs.~\ref{fig:CS12c_EuMnSb2_PPMS}(b)--(c) demonstrate that the charge transport in \ems\, is extremely sensitive to the applied field direction. The large magnetoresistance which arises when the field is aligned with the $[100]$  direction (perpendicular to the in--plane Eu magnetic moments) is suppressed as soon as there is a small component of field along the $[010]$ direction.
	
	
	The extent to which the electron bands are altered remains to be seen since ARPES cannot be performed in an applied magnetic field. Hence, a natural extension of the present work would be  an \textit{ab initio} study of the electronic band structure with 
	different spin configurations, the latter being determined experimentally by resonant elastic x-ray scattering (REXS) in an applied field. As REXS is element-specific, it allows for the study of the spin configuration on the Eu and Mn magnetic sublattices separately.

	In conclusion, we have used powder neutron diffraction to establish that \ems\, displays magnetic ordering at two distinct temperatures, and we have determined the spin arrangement for both magnetic sublattices in spite of the large neutron absorption of Eu. Moreover, in the small energy window available to map the electronic spectrum due to the presence of strongly correlated Eu $4f$ bands, we have successfully recorded ARPES spectra which provide evidence for a Dirac-like linear dispersion at $E_{\rm F}$. Finally, our magnetotransport results show that in \ems\ there is a strong coupling between magnetic order and charge transport. These three strands of evidence lend support to the prediction that \ems\,is a promising material to realize magnetic control of topological quasiparticles.
	
	\begin{acknowledgments}
		The authors wish to thank D. D. Khalyavin for technical help. This work was supported by the U.K. Engineering and Physical Sciences Research Council, Grant Nos. EP/N034872/1 and EP/M020517/1, National Key Research and Development Program of China (2017YFA0302901), the National Natural Science Foundation of China (11774399) and the Beijing Natural Science Foundation (Z180008). J.-R. Soh acknowledges support from the Singapore National Science Scholarship, Agency for Science Technology and Research.

	\end{acknowledgments}
	\nocite{*}
	\bibliography{library}

\begin{thebibliography}{40}%
\makeatletter
\providecommand \@ifxundefined [1]{%
 \@ifx{#1\undefined}
}%
\providecommand \@ifnum [1]{%
 \ifnum #1\expandafter \@firstoftwo
 \else \expandafter \@secondoftwo
 \fi
}%
\providecommand \@ifx [1]{%
 \ifx #1\expandafter \@firstoftwo
 \else \expandafter \@secondoftwo
 \fi
}%
\providecommand \natexlab [1]{#1}%
\providecommand \enquote  [1]{``#1''}%
\providecommand \bibnamefont  [1]{#1}%
\providecommand \bibfnamefont [1]{#1}%
\providecommand \citenamefont [1]{#1}%
\providecommand \href@noop [0]{\@secondoftwo}%
\providecommand \href [0]{\begingroup \@sanitize@url \@href}%
\providecommand \@href[1]{\@@startlink{#1}\@@href}%
\providecommand \@@href[1]{\endgroup#1\@@endlink}%
\providecommand \@sanitize@url [0]{\catcode `\\12\catcode `\$12\catcode
  `\&12\catcode `\#12\catcode `\^12\catcode `\_12\catcode `\%12\relax}%
\providecommand \@@startlink[1]{}%
\providecommand \@@endlink[0]{}%
\providecommand \url  [0]{\begingroup\@sanitize@url \@url }%
\providecommand \@url [1]{\endgroup\@href {#1}{\urlprefix }}%
\providecommand \urlprefix  [0]{URL }%
\providecommand \Eprint [0]{\href }%
\providecommand \doibase [0]{http://dx.doi.org/}%
\providecommand \selectlanguage [0]{\@gobble}%
\providecommand \bibinfo  [0]{\@secondoftwo}%
\providecommand \bibfield  [0]{\@secondoftwo}%
\providecommand \translation [1]{[#1]}%
\providecommand \BibitemOpen [0]{}%
\providecommand \bibitemStop [0]{}%
\providecommand \bibitemNoStop [0]{.\EOS\space}%
\providecommand \EOS [0]{\spacefactor3000\relax}%
\providecommand \BibitemShut  [1]{\csname bibitem#1\endcsname}%
\let\auto@bib@innerbib\@empty
\bibitem [{\citenamefont {Armitage}\ \emph {et~al.}(2018)\citenamefont
  {Armitage}, \citenamefont {Mele},\ and\ \citenamefont
  {Vishwanath}}]{Armitage2018}%
  \BibitemOpen
  \bibfield  {author} {\bibinfo {author} {\bibfnamefont {N.~P.}\ \bibnamefont
  {Armitage}}, \bibinfo {author} {\bibfnamefont {E.~J.}\ \bibnamefont {Mele}},
  \ and\ \bibinfo {author} {\bibfnamefont {A.}~\bibnamefont {Vishwanath}},\
  }\href {\doibase 10.1103/RevModPhys.90.015001} {\bibfield  {journal}
  {\bibinfo  {journal} {Rev. Mod. Phys.}\ }\textbf {\bibinfo {volume} {90}},\
  \bibinfo {pages} {015001} (\bibinfo {year} {2018})}\BibitemShut {NoStop}%
\bibitem [{\citenamefont {Burkov}(2016)}]{Burkov2016}%
  \BibitemOpen
  \bibfield  {author} {\bibinfo {author} {\bibfnamefont {A.~A.}\ \bibnamefont
  {Burkov}},\ }\href {\doibase 10.1038/nmat4788} {\bibfield  {journal}
  {\bibinfo  {journal} {Nat. Mat.}\ }\textbf {\bibinfo {volume} {15}},\
  \bibinfo {pages} {1145} (\bibinfo {year} {2016})}\BibitemShut {NoStop}%
\bibitem [{\citenamefont {Rau}\ \emph {et~al.}(2016)\citenamefont {Rau},
  \citenamefont {Lee},\ and\ \citenamefont {Kee}}]{Rau2016}%
  \BibitemOpen
  \bibfield  {author} {\bibinfo {author} {\bibfnamefont {J.~G.}\ \bibnamefont
  {Rau}}, \bibinfo {author} {\bibfnamefont {E.~K.-H.}\ \bibnamefont {Lee}}, \
  and\ \bibinfo {author} {\bibfnamefont {H.-Y.}\ \bibnamefont {Kee}},\ }\href
  {\doibase 10.1146/annurev-conmatphys-031115-011319} {\bibfield  {journal}
  {\bibinfo  {journal} {Ann. Rev. Con. Mat. Phys.}\ }\textbf {\bibinfo {volume}
  {7}},\ \bibinfo {pages} {195} (\bibinfo {year} {2016})}\BibitemShut {NoStop}%
\bibitem [{\citenamefont {Hasan}\ and\ \citenamefont {Kane}(2010)}]{Hasan2010}%
  \BibitemOpen
  \bibfield  {author} {\bibinfo {author} {\bibfnamefont {M.~Z.}\ \bibnamefont
  {Hasan}}\ and\ \bibinfo {author} {\bibfnamefont {C.~L.}\ \bibnamefont
  {Kane}},\ }\href {\doibase 10.1103/RevModPhys.82.3045} {\bibfield  {journal}
  {\bibinfo  {journal} {Rev. Mod. Phys.}\ }\textbf {\bibinfo {volume} {82}},\
  \bibinfo {pages} {3045} (\bibinfo {year} {2010})}\BibitemShut {NoStop}%
\bibitem [{\citenamefont {Pesin}\ and\ \citenamefont
  {Balents}(2010)}]{Pesin2009}%
  \BibitemOpen
  \bibfield  {author} {\bibinfo {author} {\bibfnamefont {D.}~\bibnamefont
  {Pesin}}\ and\ \bibinfo {author} {\bibfnamefont {L.}~\bibnamefont
  {Balents}},\ }\href {\doibase 10.1038/nphys1606} {\bibfield  {journal}
  {\bibinfo  {journal} {Nat. Phys.}\ }\textbf {\bibinfo {volume} {6}},\
  \bibinfo {pages} {376} (\bibinfo {year} {2010})}\BibitemShut {NoStop}%
\bibitem [{\citenamefont {\v{S}mejkal}\ \emph
  {et~al.}(2017{\natexlab{a}})\citenamefont {\v{S}mejkal}, \citenamefont
  {Jungwirth},\ and\ \citenamefont {Sinova}}]{mejkal2017}%
  \BibitemOpen
  \bibfield  {author} {\bibinfo {author} {\bibfnamefont {L.}~\bibnamefont
  {\v{S}mejkal}}, \bibinfo {author} {\bibfnamefont {T.}~\bibnamefont
  {Jungwirth}}, \ and\ \bibinfo {author} {\bibfnamefont {J.}~\bibnamefont
  {Sinova}},\ }\href {\doibase 10.1002/pssr.201770317} {\bibfield  {journal}
  {\bibinfo  {journal} {Phys. Status Solidi B}\ }\textbf {\bibinfo {volume}
  {11}},\ \bibinfo {pages} {1770317} (\bibinfo {year}
  {2017}{\natexlab{a}})}\BibitemShut {NoStop}%
\bibitem [{\citenamefont {\v{S}mejkal}\ \emph
  {et~al.}(2017{\natexlab{b}})\citenamefont {\v{S}mejkal}, \citenamefont
  {\v{Z}elezn\'y}, \citenamefont {Sinova},\ and\ \citenamefont
  {Jungwirth}}]{Smejkal2017}%
  \BibitemOpen
  \bibfield  {author} {\bibinfo {author} {\bibfnamefont {L.}~\bibnamefont
  {\v{S}mejkal}}, \bibinfo {author} {\bibfnamefont {J.}~\bibnamefont
  {\v{Z}elezn\'y}}, \bibinfo {author} {\bibfnamefont {J.}~\bibnamefont
  {Sinova}}, \ and\ \bibinfo {author} {\bibfnamefont {T.}~\bibnamefont
  {Jungwirth}},\ }\href {\doibase 10.1103/PhysRevLett.118.106402} {\bibfield
  {journal} {\bibinfo  {journal} {Phys. Rev. Lett.}\ }\textbf {\bibinfo
  {volume} {118}},\ \bibinfo {pages} {106402} (\bibinfo {year}
  {2017}{\natexlab{b}})}\BibitemShut {NoStop}%
\bibitem [{\citenamefont {Shi}\ \emph {et~al.}(2015)\citenamefont {Shi},
  \citenamefont {Wang},\ and\ \citenamefont {Wu}}]{Shi2015}%
  \BibitemOpen
  \bibfield  {author} {\bibinfo {author} {\bibfnamefont {Z.}~\bibnamefont
  {Shi}}, \bibinfo {author} {\bibfnamefont {M.}~\bibnamefont {Wang}}, \ and\
  \bibinfo {author} {\bibfnamefont {J.}~\bibnamefont {Wu}},\ }\href {\doibase
  10.1063/1.4930875} {\bibfield  {journal} {\bibinfo  {journal} {Appl. Phys.
  Lett.}\ }\textbf {\bibinfo {volume} {107}},\ \bibinfo {pages} {102403}
  (\bibinfo {year} {2015})}\BibitemShut {NoStop}%
\bibitem [{\citenamefont {Liu}\ \emph {et~al.}(2016)\citenamefont {Liu},
  \citenamefont {Hu}, \citenamefont {Cao}, \citenamefont {Zhu}, \citenamefont
  {Chuang}, \citenamefont {Graf}, \citenamefont {Adams}, \citenamefont
  {Radmanesh}, \citenamefont {Spinu}, \citenamefont {Chiorescu},\ and\
  \citenamefont {Mao}}]{Liu2016}%
  \BibitemOpen
  \bibfield  {author} {\bibinfo {author} {\bibfnamefont {J.~Y.}\ \bibnamefont
  {Liu}}, \bibinfo {author} {\bibfnamefont {J.}~\bibnamefont {Hu}}, \bibinfo
  {author} {\bibfnamefont {H.}~\bibnamefont {Cao}}, \bibinfo {author}
  {\bibfnamefont {Y.}~\bibnamefont {Zhu}}, \bibinfo {author} {\bibfnamefont
  {A.}~\bibnamefont {Chuang}}, \bibinfo {author} {\bibfnamefont
  {D.}~\bibnamefont {Graf}}, \bibinfo {author} {\bibfnamefont {D.~J.}\
  \bibnamefont {Adams}}, \bibinfo {author} {\bibfnamefont {S.~M.~A.}\
  \bibnamefont {Radmanesh}}, \bibinfo {author} {\bibfnamefont {L.}~\bibnamefont
  {Spinu}}, \bibinfo {author} {\bibfnamefont {I.}~\bibnamefont {Chiorescu}}, \
  and\ \bibinfo {author} {\bibfnamefont {Z.~Q.}\ \bibnamefont {Mao}},\ }\href
  {\doibase 10.1038/srep30525} {\bibfield  {journal} {\bibinfo  {journal} {Sci.
  Rep.}\ }\textbf {\bibinfo {volume} {6}},\ \bibinfo {pages} {30525} (\bibinfo
  {year} {2016})}\BibitemShut {NoStop}%
\bibitem [{\citenamefont {Farhan}\ \emph {et~al.}(2014)\citenamefont {Farhan},
  \citenamefont {Lee},\ and\ \citenamefont {Shim}}]{Farhan2014}%
  \BibitemOpen
  \bibfield  {author} {\bibinfo {author} {\bibfnamefont {M.~A.}\ \bibnamefont
  {Farhan}}, \bibinfo {author} {\bibfnamefont {G.}~\bibnamefont {Lee}}, \ and\
  \bibinfo {author} {\bibfnamefont {J.~H.}\ \bibnamefont {Shim}},\ }\href
  {\doibase 10.1088/0953-8984/26/4/042201} {\bibfield  {journal} {\bibinfo
  {journal} {J. Phys.: Condens. Matter}\ }\textbf {\bibinfo {volume} {26}},\
  \bibinfo {pages} {042201} (\bibinfo {year} {2014})}\BibitemShut {NoStop}%
\bibitem [{\citenamefont {Kealhofer}\ \emph {et~al.}(2018)\citenamefont
  {Kealhofer}, \citenamefont {Jang}, \citenamefont {Griffin}, \citenamefont
  {John}, \citenamefont {Benavides}, \citenamefont {Doyle}, \citenamefont
  {Helm}, \citenamefont {Moll}, \citenamefont {Neaton}, \citenamefont {Chan},
  \citenamefont {Denlinger},\ and\ \citenamefont {Analytis}}]{Kealhofer2018}%
  \BibitemOpen
  \bibfield  {author} {\bibinfo {author} {\bibfnamefont {R.}~\bibnamefont
  {Kealhofer}}, \bibinfo {author} {\bibfnamefont {S.}~\bibnamefont {Jang}},
  \bibinfo {author} {\bibfnamefont {S.~M.}\ \bibnamefont {Griffin}}, \bibinfo
  {author} {\bibfnamefont {C.}~\bibnamefont {John}}, \bibinfo {author}
  {\bibfnamefont {K.~A.}\ \bibnamefont {Benavides}}, \bibinfo {author}
  {\bibfnamefont {S.}~\bibnamefont {Doyle}}, \bibinfo {author} {\bibfnamefont
  {T.}~\bibnamefont {Helm}}, \bibinfo {author} {\bibfnamefont {P.~J.~W.}\
  \bibnamefont {Moll}}, \bibinfo {author} {\bibfnamefont {J.~B.}\ \bibnamefont
  {Neaton}}, \bibinfo {author} {\bibfnamefont {J.~Y.}\ \bibnamefont {Chan}},
  \bibinfo {author} {\bibfnamefont {J.~D.}\ \bibnamefont {Denlinger}}, \ and\
  \bibinfo {author} {\bibfnamefont {J.~G.}\ \bibnamefont {Analytis}},\ }\href
  {\doibase 10.1103/PhysRevB.97.045109} {\bibfield  {journal} {\bibinfo
  {journal} {Phys. Rev. B}\ }\textbf {\bibinfo {volume} {97}},\ \bibinfo
  {pages} {045109} (\bibinfo {year} {2018})}\BibitemShut {NoStop}%
\bibitem [{\citenamefont {Wang}\ \emph {et~al.}(2018)\citenamefont {Wang},
  \citenamefont {Xu}, \citenamefont {Sun},\ and\ \citenamefont
  {Xia}}]{Wang2018}%
  \BibitemOpen
  \bibfield  {author} {\bibinfo {author} {\bibfnamefont {Y.-Y.}\ \bibnamefont
  {Wang}}, \bibinfo {author} {\bibfnamefont {S.}~\bibnamefont {Xu}}, \bibinfo
  {author} {\bibfnamefont {L.-L.}\ \bibnamefont {Sun}}, \ and\ \bibinfo
  {author} {\bibfnamefont {T.-L.}\ \bibnamefont {Xia}},\ }\href {\doibase
  10.1103/PhysRevMaterials.2.021201} {\bibfield  {journal} {\bibinfo  {journal}
  {Phys. Rev. Mater.}\ }\textbf {\bibinfo {volume} {2}},\ \bibinfo {pages}
  {021201} (\bibinfo {year} {2018})}\BibitemShut {NoStop}%
\bibitem [{\citenamefont {Ramankutty}\ \emph {et~al.}(2018)\citenamefont
  {Ramankutty}, \citenamefont {Henke}, \citenamefont {Schiphorst},
  \citenamefont {Nutakki}, \citenamefont {Bron}, \citenamefont
  {Araizi-Kanoutas}, \citenamefont {Mishra}, \citenamefont {Li}, \citenamefont
  {Huang}, \citenamefont {Kim}, \citenamefont {Hoesch}, \citenamefont
  {Schlueter}, \citenamefont {Lee}, \citenamefont {de~Visser}, \citenamefont
  {Zhong}, \citenamefont {van Wezel}, \citenamefont {van Heumen},\ and\
  \citenamefont {Golden}}]{Ramankutty2018}%
  \BibitemOpen
  \bibfield  {author} {\bibinfo {author} {\bibfnamefont {S.~V.}\ \bibnamefont
  {Ramankutty}}, \bibinfo {author} {\bibfnamefont {J.}~\bibnamefont {Henke}},
  \bibinfo {author} {\bibfnamefont {A.}~\bibnamefont {Schiphorst}}, \bibinfo
  {author} {\bibfnamefont {R.}~\bibnamefont {Nutakki}}, \bibinfo {author}
  {\bibfnamefont {S.}~\bibnamefont {Bron}}, \bibinfo {author} {\bibfnamefont
  {G.}~\bibnamefont {Araizi-Kanoutas}}, \bibinfo {author} {\bibfnamefont
  {S.~K.}\ \bibnamefont {Mishra}}, \bibinfo {author} {\bibfnamefont
  {L.}~\bibnamefont {Li}}, \bibinfo {author} {\bibfnamefont {Y.~K.}\
  \bibnamefont {Huang}}, \bibinfo {author} {\bibfnamefont {T.~K.}\ \bibnamefont
  {Kim}}, \bibinfo {author} {\bibfnamefont {M.}~\bibnamefont {Hoesch}},
  \bibinfo {author} {\bibfnamefont {C.}~\bibnamefont {Schlueter}}, \bibinfo
  {author} {\bibfnamefont {T.~L.}\ \bibnamefont {Lee}}, \bibinfo {author}
  {\bibfnamefont {A.}~\bibnamefont {de~Visser}}, \bibinfo {author}
  {\bibfnamefont {Z.}~\bibnamefont {Zhong}}, \bibinfo {author} {\bibfnamefont
  {J.}~\bibnamefont {van Wezel}}, \bibinfo {author} {\bibfnamefont
  {E.}~\bibnamefont {van Heumen}}, \ and\ \bibinfo {author} {\bibfnamefont
  {M.~S.}\ \bibnamefont {Golden}},\ }\href {\doibase
  10.21468/SciPostPhys.4.2.010} {\bibfield  {journal} {\bibinfo  {journal}
  {SciPost Phys.}\ }\textbf {\bibinfo {volume} {4}},\ \bibinfo {pages} {010}
  (\bibinfo {year} {2018})}\BibitemShut {NoStop}%
\bibitem [{\citenamefont {Liu}\ \emph {et~al.}(2017{\natexlab{a}})\citenamefont
  {Liu}, \citenamefont {Hu}, \citenamefont {Zhang}, \citenamefont {Graf},
  \citenamefont {Cao}, \citenamefont {Radmanesh}, \citenamefont {Adams},
  \citenamefont {Zhu}, \citenamefont {Cheng}, \citenamefont {Liu},
  \citenamefont {Phelan}, \citenamefont {Wei}, \citenamefont {Jaime},
  \citenamefont {Balakirev}, \citenamefont {Tennant}, \citenamefont {DiTusa},
  \citenamefont {Chiorescu}, \citenamefont {Spinu},\ and\ \citenamefont
  {Mao}}]{Liu2017b}%
  \BibitemOpen
  \bibfield  {author} {\bibinfo {author} {\bibfnamefont {J.~Y.}\ \bibnamefont
  {Liu}}, \bibinfo {author} {\bibfnamefont {J.}~\bibnamefont {Hu}}, \bibinfo
  {author} {\bibfnamefont {Q.}~\bibnamefont {Zhang}}, \bibinfo {author}
  {\bibfnamefont {D.}~\bibnamefont {Graf}}, \bibinfo {author} {\bibfnamefont
  {H.~B.}\ \bibnamefont {Cao}}, \bibinfo {author} {\bibfnamefont {S.~M.~A.}\
  \bibnamefont {Radmanesh}}, \bibinfo {author} {\bibfnamefont {D.~J.}\
  \bibnamefont {Adams}}, \bibinfo {author} {\bibfnamefont {Y.~L.}\ \bibnamefont
  {Zhu}}, \bibinfo {author} {\bibfnamefont {G.}~\bibnamefont {Cheng}}, \bibinfo
  {author} {\bibfnamefont {X.}~\bibnamefont {Liu}}, \bibinfo {author}
  {\bibfnamefont {W.~A.}\ \bibnamefont {Phelan}}, \bibinfo {author}
  {\bibfnamefont {J.}~\bibnamefont {Wei}}, \bibinfo {author} {\bibfnamefont
  {M.}~\bibnamefont {Jaime}}, \bibinfo {author} {\bibfnamefont
  {F.}~\bibnamefont {Balakirev}}, \bibinfo {author} {\bibfnamefont {D.~A.}\
  \bibnamefont {Tennant}}, \bibinfo {author} {\bibfnamefont {J.~F.}\
  \bibnamefont {DiTusa}}, \bibinfo {author} {\bibfnamefont {I.}~\bibnamefont
  {Chiorescu}}, \bibinfo {author} {\bibfnamefont {L.}~\bibnamefont {Spinu}}, \
  and\ \bibinfo {author} {\bibfnamefont {Z.~Q.}\ \bibnamefont {Mao}},\ }\href
  {\doibase 10.1038/nmat4953} {\bibfield  {journal} {\bibinfo  {journal} {Nat.
  Mat.}\ }\textbf {\bibinfo {volume} {16}},\ \bibinfo {pages} {905} (\bibinfo
  {year} {2017}{\natexlab{a}})}\BibitemShut {NoStop}%
\bibitem [{\citenamefont {He}\ \emph {et~al.}(2017)\citenamefont {He},
  \citenamefont {Fu}, \citenamefont {Zhao}, \citenamefont {Liang},
  \citenamefont {Chen}, \citenamefont {Leng}, \citenamefont {Wang},
  \citenamefont {Li}, \citenamefont {Zhang}, \citenamefont {Xue}, \citenamefont
  {Li}, \citenamefont {Zhang}, \citenamefont {Ren},\ and\ \citenamefont
  {Chen}}]{He2017}%
  \BibitemOpen
  \bibfield  {author} {\bibinfo {author} {\bibfnamefont {J.~B.}\ \bibnamefont
  {He}}, \bibinfo {author} {\bibfnamefont {Y.}~\bibnamefont {Fu}}, \bibinfo
  {author} {\bibfnamefont {L.~X.}\ \bibnamefont {Zhao}}, \bibinfo {author}
  {\bibfnamefont {H.}~\bibnamefont {Liang}}, \bibinfo {author} {\bibfnamefont
  {D.}~\bibnamefont {Chen}}, \bibinfo {author} {\bibfnamefont {Y.~M.}\
  \bibnamefont {Leng}}, \bibinfo {author} {\bibfnamefont {X.~M.}\ \bibnamefont
  {Wang}}, \bibinfo {author} {\bibfnamefont {J.}~\bibnamefont {Li}}, \bibinfo
  {author} {\bibfnamefont {S.}~\bibnamefont {Zhang}}, \bibinfo {author}
  {\bibfnamefont {M.~Q.}\ \bibnamefont {Xue}}, \bibinfo {author} {\bibfnamefont
  {C.~H.}\ \bibnamefont {Li}}, \bibinfo {author} {\bibfnamefont
  {P.}~\bibnamefont {Zhang}}, \bibinfo {author} {\bibfnamefont {Z.~A.}\
  \bibnamefont {Ren}}, \ and\ \bibinfo {author} {\bibfnamefont {G.~F.}\
  \bibnamefont {Chen}},\ }\href {\doibase 10.1103/PhysRevB.95.045128}
  {\bibfield  {journal} {\bibinfo  {journal} {Phys. Rev. B}\ }\textbf {\bibinfo
  {volume} {95}},\ \bibinfo {pages} {045128} (\bibinfo {year}
  {2017})}\BibitemShut {NoStop}%
\bibitem [{\citenamefont {You}\ \emph {et~al.}(2019)\citenamefont {You},
  \citenamefont {Lee}, \citenamefont {Choi}, \citenamefont {Jo}, \citenamefont
  {Shim},\ and\ \citenamefont {Kim}}]{You2019}%
  \BibitemOpen
  \bibfield  {author} {\bibinfo {author} {\bibfnamefont {J.~S.}\ \bibnamefont
  {You}}, \bibinfo {author} {\bibfnamefont {I.}~\bibnamefont {Lee}}, \bibinfo
  {author} {\bibfnamefont {E.}~\bibnamefont {Choi}}, \bibinfo {author}
  {\bibfnamefont {Y.}~\bibnamefont {Jo}}, \bibinfo {author} {\bibfnamefont
  {J.}~\bibnamefont {Shim}}, \ and\ \bibinfo {author} {\bibfnamefont {J.~S.}\
  \bibnamefont {Kim}},\ }\href {\doibase
  https://doi.org/10.1016/j.cap.2018.10.022} {\bibfield  {journal} {\bibinfo
  {journal} {Curr. Appl. Phys.}\ }\textbf {\bibinfo {volume} {19}},\ \bibinfo
  {pages} {230 } (\bibinfo {year} {2019})}\BibitemShut {NoStop}%
\bibitem [{\citenamefont {Yan}\ \emph {et~al.}(2017)\citenamefont {Yan},
  \citenamefont {Zhang}, \citenamefont {Liu}, \citenamefont {Liu},
  \citenamefont {Zhang}, \citenamefont {Xiu},\ and\ \citenamefont
  {Zhou}}]{Yan2017}%
  \BibitemOpen
  \bibfield  {author} {\bibinfo {author} {\bibfnamefont {X.}~\bibnamefont
  {Yan}}, \bibinfo {author} {\bibfnamefont {C.}~\bibnamefont {Zhang}}, \bibinfo
  {author} {\bibfnamefont {S.-S.}\ \bibnamefont {Liu}}, \bibinfo {author}
  {\bibfnamefont {Y.-W.}\ \bibnamefont {Liu}}, \bibinfo {author} {\bibfnamefont
  {D.~W.}\ \bibnamefont {Zhang}}, \bibinfo {author} {\bibfnamefont {F.-X.}\
  \bibnamefont {Xiu}}, \ and\ \bibinfo {author} {\bibfnamefont
  {P.}~\bibnamefont {Zhou}},\ }\href {\doibase 10.1007/s11467-017-0663-0}
  {\bibfield  {journal} {\bibinfo  {journal} {Front. Phys.}\ }\textbf {\bibinfo
  {volume} {12}},\ \bibinfo {pages} {127209} (\bibinfo {year}
  {2017})}\BibitemShut {NoStop}%
\bibitem [{\citenamefont {Ray}\ and\ \citenamefont {Alff}(2017)}]{Soumya2017}%
  \BibitemOpen
  \bibfield  {author} {\bibinfo {author} {\bibfnamefont {S.~J.}\ \bibnamefont
  {Ray}}\ and\ \bibinfo {author} {\bibfnamefont {L.}~\bibnamefont {Alff}},\
  }\href {\doibase 10.1002/pssb.201600163} {\bibfield  {journal} {\bibinfo
  {journal} {Phys. Status Solidi B}\ }\textbf {\bibinfo {volume} {254}},\
  \bibinfo {pages} {1600163} (\bibinfo {year} {2017})}\BibitemShut {NoStop}%
\bibitem [{\citenamefont {Klemenz}\ \emph {et~al.}(2019)\citenamefont
  {Klemenz}, \citenamefont {Lei},\ and\ \citenamefont {Schoop}}]{Klemenz2019}%
  \BibitemOpen
  \bibfield  {author} {\bibinfo {author} {\bibfnamefont {S.}~\bibnamefont
  {Klemenz}}, \bibinfo {author} {\bibfnamefont {S.}~\bibnamefont {Lei}}, \ and\
  \bibinfo {author} {\bibfnamefont {L.~M.}\ \bibnamefont {Schoop}},\ }\href
  {\doibase 10.1146/annurev-matsci-070218-010114} {\bibfield  {journal}
  {\bibinfo  {journal} {Annu. Rev. Mater. Res.}\ }\textbf {\bibinfo {volume}
  {49}},\ \bibinfo {pages} {null} (\bibinfo {year} {2019})}\BibitemShut
  {NoStop}%
\bibitem [{\citenamefont {Yi}\ \emph {et~al.}(2017)\citenamefont {Yi},
  \citenamefont {Yang}, \citenamefont {Yang}, \citenamefont {Wang},
  \citenamefont {Matsushita}, \citenamefont {Miao}, \citenamefont {Jiao},
  \citenamefont {Cheng}, \citenamefont {Li}, \citenamefont {Yamaura},
  \citenamefont {Shi},\ and\ \citenamefont {Luo}}]{Yi2017}%
  \BibitemOpen
  \bibfield  {author} {\bibinfo {author} {\bibfnamefont {C.}~\bibnamefont
  {Yi}}, \bibinfo {author} {\bibfnamefont {S.}~\bibnamefont {Yang}}, \bibinfo
  {author} {\bibfnamefont {M.}~\bibnamefont {Yang}}, \bibinfo {author}
  {\bibfnamefont {L.}~\bibnamefont {Wang}}, \bibinfo {author} {\bibfnamefont
  {Y.}~\bibnamefont {Matsushita}}, \bibinfo {author} {\bibfnamefont
  {S.}~\bibnamefont {Miao}}, \bibinfo {author} {\bibfnamefont {Y.}~\bibnamefont
  {Jiao}}, \bibinfo {author} {\bibfnamefont {J.}~\bibnamefont {Cheng}},
  \bibinfo {author} {\bibfnamefont {Y.}~\bibnamefont {Li}}, \bibinfo {author}
  {\bibfnamefont {K.}~\bibnamefont {Yamaura}}, \bibinfo {author} {\bibfnamefont
  {Y.}~\bibnamefont {Shi}}, \ and\ \bibinfo {author} {\bibfnamefont
  {J.}~\bibnamefont {Luo}},\ }\href {\doibase 10.1103/PhysRevB.96.205103}
  {\bibfield  {journal} {\bibinfo  {journal} {Phys. Rev. B}\ }\textbf {\bibinfo
  {volume} {96}},\ \bibinfo {pages} {205103} (\bibinfo {year}
  {2017})}\BibitemShut {NoStop}%
\bibitem [{\citenamefont {Wang}\ \emph {et~al.}(2012)\citenamefont {Wang},
  \citenamefont {Graf}, \citenamefont {Wang}, \citenamefont {Lei},
  \citenamefont {Tozer},\ and\ \citenamefont {Petrovic}}]{Wang2012}%
  \BibitemOpen
  \bibfield  {author} {\bibinfo {author} {\bibfnamefont {K.}~\bibnamefont
  {Wang}}, \bibinfo {author} {\bibfnamefont {D.}~\bibnamefont {Graf}}, \bibinfo
  {author} {\bibfnamefont {L.}~\bibnamefont {Wang}}, \bibinfo {author}
  {\bibfnamefont {H.}~\bibnamefont {Lei}}, \bibinfo {author} {\bibfnamefont
  {S.~W.}\ \bibnamefont {Tozer}}, \ and\ \bibinfo {author} {\bibfnamefont
  {C.}~\bibnamefont {Petrovic}},\ }\href {\doibase 10.1103/PhysRevB.85.041101}
  {\bibfield  {journal} {\bibinfo  {journal} {Phys. Rev. B}\ }\textbf {\bibinfo
  {volume} {85}},\ \bibinfo {pages} {041101(R)} (\bibinfo {year}
  {2012})}\BibitemShut {NoStop}%
\bibitem [{\citenamefont {Wang}\ \emph {et~al.}(2016)\citenamefont {Wang},
  \citenamefont {Zaliznyak}, \citenamefont {Ren}, \citenamefont {Wu},
  \citenamefont {Graf}, \citenamefont {Garlea}, \citenamefont {Warren},
  \citenamefont {Bozin}, \citenamefont {Zhu},\ and\ \citenamefont
  {Petrovic}}]{Wang2016}%
  \BibitemOpen
  \bibfield  {author} {\bibinfo {author} {\bibfnamefont {A.}~\bibnamefont
  {Wang}}, \bibinfo {author} {\bibfnamefont {I.}~\bibnamefont {Zaliznyak}},
  \bibinfo {author} {\bibfnamefont {W.}~\bibnamefont {Ren}}, \bibinfo {author}
  {\bibfnamefont {L.}~\bibnamefont {Wu}}, \bibinfo {author} {\bibfnamefont
  {D.}~\bibnamefont {Graf}}, \bibinfo {author} {\bibfnamefont {V.~O.}\
  \bibnamefont {Garlea}}, \bibinfo {author} {\bibfnamefont {J.~B.}\
  \bibnamefont {Warren}}, \bibinfo {author} {\bibfnamefont {E.}~\bibnamefont
  {Bozin}}, \bibinfo {author} {\bibfnamefont {Y.}~\bibnamefont {Zhu}}, \ and\
  \bibinfo {author} {\bibfnamefont {C.}~\bibnamefont {Petrovic}},\ }\href
  {\doibase 10.1103/PhysRevB.94.165161} {\bibfield  {journal} {\bibinfo
  {journal} {Phys. Rev. B}\ }\textbf {\bibinfo {volume} {94}},\ \bibinfo
  {pages} {165161} (\bibinfo {year} {2016})}\BibitemShut {NoStop}%
\bibitem [{\citenamefont {Brechtel}\ \emph {et~al.}(1981)\citenamefont
  {Brechtel}, \citenamefont {Cordier},\ and\ \citenamefont
  {Schäfer}}]{Brechtel1981}%
  \BibitemOpen
  \bibfield  {author} {\bibinfo {author} {\bibfnamefont {E.}~\bibnamefont
  {Brechtel}}, \bibinfo {author} {\bibfnamefont {G.}~\bibnamefont {Cordier}}, \
  and\ \bibinfo {author} {\bibfnamefont {H.}~\bibnamefont {Schäfer}},\ }\href
  {\doibase https://doi.org/10.1016/0022-5088(81)90057-6} {\bibfield  {journal}
  {\bibinfo  {journal} {J. Less Comm. Met.}\ }\textbf {\bibinfo {volume}
  {79}},\ \bibinfo {pages} {131 } (\bibinfo {year} {1981})}\BibitemShut
  {NoStop}%
\bibitem [{\citenamefont {Park}\ \emph {et~al.}(2011)\citenamefont {Park},
  \citenamefont {Lee}, \citenamefont {Wolff-Fabris}, \citenamefont {Koh},
  \citenamefont {Eom}, \citenamefont {Kim}, \citenamefont {Farhan},
  \citenamefont {Jo}, \citenamefont {Kim}, \citenamefont {Shim},\ and\
  \citenamefont {Kim}}]{Park2011}%
  \BibitemOpen
  \bibfield  {author} {\bibinfo {author} {\bibfnamefont {J.}~\bibnamefont
  {Park}}, \bibinfo {author} {\bibfnamefont {G.}~\bibnamefont {Lee}}, \bibinfo
  {author} {\bibfnamefont {F.}~\bibnamefont {Wolff-Fabris}}, \bibinfo {author}
  {\bibfnamefont {Y.~Y.}\ \bibnamefont {Koh}}, \bibinfo {author} {\bibfnamefont
  {M.~J.}\ \bibnamefont {Eom}}, \bibinfo {author} {\bibfnamefont {Y.~K.}\
  \bibnamefont {Kim}}, \bibinfo {author} {\bibfnamefont {M.~A.}\ \bibnamefont
  {Farhan}}, \bibinfo {author} {\bibfnamefont {Y.~J.}\ \bibnamefont {Jo}},
  \bibinfo {author} {\bibfnamefont {C.}~\bibnamefont {Kim}}, \bibinfo {author}
  {\bibfnamefont {J.~H.}\ \bibnamefont {Shim}}, \ and\ \bibinfo {author}
  {\bibfnamefont {J.~S.}\ \bibnamefont {Kim}},\ }\href {\doibase
  10.1103/PhysRevLett.107.126402} {\bibfield  {journal} {\bibinfo  {journal}
  {Phys. Rev. Lett.}\ }\textbf {\bibinfo {volume} {107}},\ \bibinfo {pages}
  {126402} (\bibinfo {year} {2011})}\BibitemShut {NoStop}%
\bibitem [{\citenamefont {Feng}\ \emph {et~al.}(2014)\citenamefont {Feng},
  \citenamefont {Wang}, \citenamefont {Chen}, \citenamefont {Shi},
  \citenamefont {Xie}, \citenamefont {Yi}, \citenamefont {Liang}, \citenamefont
  {He}, \citenamefont {He}, \citenamefont {Peng}, \citenamefont {Liu},
  \citenamefont {Liu}, \citenamefont {Zhao}, \citenamefont {Liu}, \citenamefont
  {Dong}, \citenamefont {Zhang}, \citenamefont {Chen}, \citenamefont {Xu},
  \citenamefont {Dai}, \citenamefont {Fang},\ and\ \citenamefont
  {Zhou}}]{Feng2014}%
  \BibitemOpen
  \bibfield  {author} {\bibinfo {author} {\bibfnamefont {Y.}~\bibnamefont
  {Feng}}, \bibinfo {author} {\bibfnamefont {Z.}~\bibnamefont {Wang}}, \bibinfo
  {author} {\bibfnamefont {C.}~\bibnamefont {Chen}}, \bibinfo {author}
  {\bibfnamefont {Y.}~\bibnamefont {Shi}}, \bibinfo {author} {\bibfnamefont
  {Z.}~\bibnamefont {Xie}}, \bibinfo {author} {\bibfnamefont {H.}~\bibnamefont
  {Yi}}, \bibinfo {author} {\bibfnamefont {A.}~\bibnamefont {Liang}}, \bibinfo
  {author} {\bibfnamefont {S.}~\bibnamefont {He}}, \bibinfo {author}
  {\bibfnamefont {J.}~\bibnamefont {He}}, \bibinfo {author} {\bibfnamefont
  {Y.}~\bibnamefont {Peng}}, \bibinfo {author} {\bibfnamefont {X.}~\bibnamefont
  {Liu}}, \bibinfo {author} {\bibfnamefont {Y.}~\bibnamefont {Liu}}, \bibinfo
  {author} {\bibfnamefont {L.}~\bibnamefont {Zhao}}, \bibinfo {author}
  {\bibfnamefont {G.}~\bibnamefont {Liu}}, \bibinfo {author} {\bibfnamefont
  {X.}~\bibnamefont {Dong}}, \bibinfo {author} {\bibfnamefont {J.}~\bibnamefont
  {Zhang}}, \bibinfo {author} {\bibfnamefont {C.}~\bibnamefont {Chen}},
  \bibinfo {author} {\bibfnamefont {Z.}~\bibnamefont {Xu}}, \bibinfo {author}
  {\bibfnamefont {X.}~\bibnamefont {Dai}}, \bibinfo {author} {\bibfnamefont
  {Z.}~\bibnamefont {Fang}}, \ and\ \bibinfo {author} {\bibfnamefont {X.~J.}\
  \bibnamefont {Zhou}},\ }\href {\doibase doi:10.1038/srep05385} {\bibfield
  {journal} {\bibinfo  {journal} {Sci, Rep.}\ }\textbf {\bibinfo {volume}
  {4}},\ \bibinfo {pages} {5385} (\bibinfo {year} {2014})}\BibitemShut
  {NoStop}%
\bibitem [{\citenamefont {Masuda}\ \emph {et~al.}(2016)\citenamefont {Masuda},
  \citenamefont {Sakai}, \citenamefont {Tokunaga}, \citenamefont {Yamasaki},
  \citenamefont {Miyake}, \citenamefont {Shiogai}, \citenamefont {Nakamura},
  \citenamefont {Awaji}, \citenamefont {Tsukazaki}, \citenamefont {Nakao},
  \citenamefont {Murakami}, \citenamefont {Arima}, \citenamefont {Tokura},\
  and\ \citenamefont {Ishiwata}}]{Masuda2016}%
  \BibitemOpen
  \bibfield  {author} {\bibinfo {author} {\bibfnamefont {H.}~\bibnamefont
  {Masuda}}, \bibinfo {author} {\bibfnamefont {H.}~\bibnamefont {Sakai}},
  \bibinfo {author} {\bibfnamefont {M.}~\bibnamefont {Tokunaga}}, \bibinfo
  {author} {\bibfnamefont {Y.}~\bibnamefont {Yamasaki}}, \bibinfo {author}
  {\bibfnamefont {A.}~\bibnamefont {Miyake}}, \bibinfo {author} {\bibfnamefont
  {J.}~\bibnamefont {Shiogai}}, \bibinfo {author} {\bibfnamefont
  {S.}~\bibnamefont {Nakamura}}, \bibinfo {author} {\bibfnamefont
  {S.}~\bibnamefont {Awaji}}, \bibinfo {author} {\bibfnamefont
  {A.}~\bibnamefont {Tsukazaki}}, \bibinfo {author} {\bibfnamefont
  {H.}~\bibnamefont {Nakao}}, \bibinfo {author} {\bibfnamefont
  {Y.}~\bibnamefont {Murakami}}, \bibinfo {author} {\bibfnamefont {T.-h.}\
  \bibnamefont {Arima}}, \bibinfo {author} {\bibfnamefont {Y.}~\bibnamefont
  {Tokura}}, \ and\ \bibinfo {author} {\bibfnamefont {S.}~\bibnamefont
  {Ishiwata}},\ }\href {http://advances.sciencemag.org/content/2/1/e1501117}
  {\bibfield  {journal} {\bibinfo  {journal} {Sci. Adv.}\ }\textbf {\bibinfo
  {volume} {2}} (\bibinfo {year} {2016})}\BibitemShut {NoStop}%
\bibitem [{\citenamefont {Jo}\ \emph {et~al.}(2014)\citenamefont {Jo},
  \citenamefont {Park}, \citenamefont {Lee}, \citenamefont {Eom}, \citenamefont
  {Choi}, \citenamefont {Shim}, \citenamefont {Kang},\ and\ \citenamefont
  {Kim}}]{Jo2014}%
  \BibitemOpen
  \bibfield  {author} {\bibinfo {author} {\bibfnamefont {Y.~J.}\ \bibnamefont
  {Jo}}, \bibinfo {author} {\bibfnamefont {J.}~\bibnamefont {Park}}, \bibinfo
  {author} {\bibfnamefont {G.}~\bibnamefont {Lee}}, \bibinfo {author}
  {\bibfnamefont {M.~J.}\ \bibnamefont {Eom}}, \bibinfo {author} {\bibfnamefont
  {E.~S.}\ \bibnamefont {Choi}}, \bibinfo {author} {\bibfnamefont {J.~H.}\
  \bibnamefont {Shim}}, \bibinfo {author} {\bibfnamefont {W.}~\bibnamefont
  {Kang}}, \ and\ \bibinfo {author} {\bibfnamefont {J.~S.}\ \bibnamefont
  {Kim}},\ }\href {\doibase 10.1103/PhysRevLett.113.156602} {\bibfield
  {journal} {\bibinfo  {journal} {Phys. Rev. Lett.}\ }\textbf {\bibinfo
  {volume} {113}},\ \bibinfo {pages} {156602} (\bibinfo {year}
  {2014})}\BibitemShut {NoStop}%
\bibitem [{\citenamefont {{Borisenko}}\ \emph {et~al.}(2015)\citenamefont
  {{Borisenko}}, \citenamefont {{Evtushinsky}}, \citenamefont {{Gibson}},
  \citenamefont {{Yaresko}}, \citenamefont {{Kim}}, \citenamefont {{Ali}},
  \citenamefont {{Buechner}}, \citenamefont {{Hoesch}},\ and\ \citenamefont
  {{Cava}}}]{Borisenko2015}%
  \BibitemOpen
  \bibfield  {author} {\bibinfo {author} {\bibfnamefont {S.}~\bibnamefont
  {{Borisenko}}}, \bibinfo {author} {\bibfnamefont {D.}~\bibnamefont
  {{Evtushinsky}}}, \bibinfo {author} {\bibfnamefont {Q.}~\bibnamefont
  {{Gibson}}}, \bibinfo {author} {\bibfnamefont {A.}~\bibnamefont {{Yaresko}}},
  \bibinfo {author} {\bibfnamefont {T.}~\bibnamefont {{Kim}}}, \bibinfo
  {author} {\bibfnamefont {M.~N.}\ \bibnamefont {{Ali}}}, \bibinfo {author}
  {\bibfnamefont {B.}~\bibnamefont {{Buechner}}}, \bibinfo {author}
  {\bibfnamefont {M.}~\bibnamefont {{Hoesch}}}, \ and\ \bibinfo {author}
  {\bibfnamefont {R.~J.}\ \bibnamefont {{Cava}}},\ }\href@noop {} {\  (\bibinfo
  {year} {2015})},\ \Eprint {http://arxiv.org/abs/arXiv:1507.04847}
  {arXiv:1507.04847} \BibitemShut {NoStop}%
\bibitem [{\citenamefont {Liu}\ \emph {et~al.}(2017{\natexlab{b}})\citenamefont
  {Liu}, \citenamefont {Hu}, \citenamefont {Graf}, \citenamefont {Zou},
  \citenamefont {Zhu}, \citenamefont {Shi}, \citenamefont {Che}, \citenamefont
  {Radmanesh}, \citenamefont {Lau}, \citenamefont {Spinu}, \citenamefont {Cao},
  \citenamefont {Ke},\ and\ \citenamefont {Mao}}]{Liu2017}%
  \BibitemOpen
  \bibfield  {author} {\bibinfo {author} {\bibfnamefont {J.~Y.}\ \bibnamefont
  {Liu}}, \bibinfo {author} {\bibfnamefont {J.}~\bibnamefont {Hu}}, \bibinfo
  {author} {\bibfnamefont {D.}~\bibnamefont {Graf}}, \bibinfo {author}
  {\bibfnamefont {T.}~\bibnamefont {Zou}}, \bibinfo {author} {\bibfnamefont
  {M.}~\bibnamefont {Zhu}}, \bibinfo {author} {\bibfnamefont {Y.}~\bibnamefont
  {Shi}}, \bibinfo {author} {\bibfnamefont {S.}~\bibnamefont {Che}}, \bibinfo
  {author} {\bibfnamefont {S.~M.~A.}\ \bibnamefont {Radmanesh}}, \bibinfo
  {author} {\bibfnamefont {C.~N.}\ \bibnamefont {Lau}}, \bibinfo {author}
  {\bibfnamefont {L.}~\bibnamefont {Spinu}}, \bibinfo {author} {\bibfnamefont
  {H.~B.}\ \bibnamefont {Cao}}, \bibinfo {author} {\bibfnamefont
  {X.}~\bibnamefont {Ke}}, \ and\ \bibinfo {author} {\bibfnamefont {Z.~Q.}\
  \bibnamefont {Mao}},\ }\href {\doibase 10.1038/s41467-017-00673-7} {\bibfield
   {journal} {\bibinfo  {journal} {Nat. Comms.}\ }\textbf {\bibinfo {volume}
  {8}},\ \bibinfo {pages} {646} (\bibinfo {year}
  {2017}{\natexlab{b}})}\BibitemShut {NoStop}%
\bibitem [{\citenamefont {May}\ \emph {et~al.}(2014)\citenamefont {May},
  \citenamefont {McGuire},\ and\ \citenamefont {Sales}}]{May2014}%
  \BibitemOpen
  \bibfield  {author} {\bibinfo {author} {\bibfnamefont {A.~F.}\ \bibnamefont
  {May}}, \bibinfo {author} {\bibfnamefont {M.~A.}\ \bibnamefont {McGuire}}, \
  and\ \bibinfo {author} {\bibfnamefont {B.~C.}\ \bibnamefont {Sales}},\ }\href
  {\doibase 10.1103/PhysRevB.90.075109} {\bibfield  {journal} {\bibinfo
  {journal} {Phys. Rev. B}\ }\textbf {\bibinfo {volume} {90}},\ \bibinfo
  {pages} {075109} (\bibinfo {year} {2014})}\BibitemShut {NoStop}%
\bibitem [{\citenamefont {Lee}\ \emph {et~al.}(2013)\citenamefont {Lee},
  \citenamefont {Farhan}, \citenamefont {Kim},\ and\ \citenamefont
  {Shim}}]{Lee2013}%
  \BibitemOpen
  \bibfield  {author} {\bibinfo {author} {\bibfnamefont {G.}~\bibnamefont
  {Lee}}, \bibinfo {author} {\bibfnamefont {M.~A.}\ \bibnamefont {Farhan}},
  \bibinfo {author} {\bibfnamefont {J.~S.}\ \bibnamefont {Kim}}, \ and\
  \bibinfo {author} {\bibfnamefont {J.~H.}\ \bibnamefont {Shim}},\ }\href
  {\doibase 10.1103/PhysRevB.87.245104} {\bibfield  {journal} {\bibinfo
  {journal} {Phys. Rev. B}\ }\textbf {\bibinfo {volume} {87}},\ \bibinfo
  {pages} {245104} (\bibinfo {year} {2013})}\BibitemShut {NoStop}%
\bibitem [{\citenamefont {Guo}\ \emph {et~al.}(2014)\citenamefont {Guo},
  \citenamefont {Princep}, \citenamefont {Zhang}, \citenamefont {Manuel},
  \citenamefont {Khalyavin}, \citenamefont {Mazin}, \citenamefont {Shi},\ and\
  \citenamefont {Boothroyd}}]{Guo2014}%
  \BibitemOpen
  \bibfield  {author} {\bibinfo {author} {\bibfnamefont {Y.~F.}\ \bibnamefont
  {Guo}}, \bibinfo {author} {\bibfnamefont {A.~J.}\ \bibnamefont {Princep}},
  \bibinfo {author} {\bibfnamefont {X.}~\bibnamefont {Zhang}}, \bibinfo
  {author} {\bibfnamefont {P.}~\bibnamefont {Manuel}}, \bibinfo {author}
  {\bibfnamefont {D.}~\bibnamefont {Khalyavin}}, \bibinfo {author}
  {\bibfnamefont {I.~I.}\ \bibnamefont {Mazin}}, \bibinfo {author}
  {\bibfnamefont {Y.~G.}\ \bibnamefont {Shi}}, \ and\ \bibinfo {author}
  {\bibfnamefont {A.~T.}\ \bibnamefont {Boothroyd}},\ }\href {\doibase
  10.1103/PhysRevB.90.075120} {\bibfield  {journal} {\bibinfo  {journal} {Phys.
  Rev. B}\ }\textbf {\bibinfo {volume} {90}},\ \bibinfo {pages} {075120}
  (\bibinfo {year} {2014})}\BibitemShut {NoStop}%
\bibitem [{\citenamefont {Chinotti}\ \emph {et~al.}(2016)\citenamefont
  {Chinotti}, \citenamefont {Pal}, \citenamefont {Ren}, \citenamefont
  {Petrovic},\ and\ \citenamefont {Degiorgi}}]{Chinotti2016}%
  \BibitemOpen
  \bibfield  {author} {\bibinfo {author} {\bibfnamefont {M.}~\bibnamefont
  {Chinotti}}, \bibinfo {author} {\bibfnamefont {A.}~\bibnamefont {Pal}},
  \bibinfo {author} {\bibfnamefont {W.~J.}\ \bibnamefont {Ren}}, \bibinfo
  {author} {\bibfnamefont {C.}~\bibnamefont {Petrovic}}, \ and\ \bibinfo
  {author} {\bibfnamefont {L.}~\bibnamefont {Degiorgi}},\ }\href {\doibase
  10.1103/PhysRevB.94.245101} {\bibfield  {journal} {\bibinfo  {journal} {Phys.
  Rev. B}\ }\textbf {\bibinfo {volume} {94}},\ \bibinfo {pages} {245101}
  (\bibinfo {year} {2016})}\BibitemShut {NoStop}%
\bibitem [{\citenamefont {Wang}\ \emph {et~al.}(2011)\citenamefont {Wang},
  \citenamefont {Zhao}, \citenamefont {Yin}, \citenamefont {Kotliar},
  \citenamefont {Kim}, \citenamefont {Aronson},\ and\ \citenamefont
  {Morosan}}]{Wang2011}%
  \BibitemOpen
  \bibfield  {author} {\bibinfo {author} {\bibfnamefont {J.~K.}\ \bibnamefont
  {Wang}}, \bibinfo {author} {\bibfnamefont {L.~L.}\ \bibnamefont {Zhao}},
  \bibinfo {author} {\bibfnamefont {Q.}~\bibnamefont {Yin}}, \bibinfo {author}
  {\bibfnamefont {G.}~\bibnamefont {Kotliar}}, \bibinfo {author} {\bibfnamefont
  {M.~S.}\ \bibnamefont {Kim}}, \bibinfo {author} {\bibfnamefont {M.~C.}\
  \bibnamefont {Aronson}}, \ and\ \bibinfo {author} {\bibfnamefont
  {E.}~\bibnamefont {Morosan}},\ }\href {\doibase 10.1103/PhysRevB.84.064428}
  {\bibfield  {journal} {\bibinfo  {journal} {Phys. Rev. B}\ }\textbf {\bibinfo
  {volume} {84}},\ \bibinfo {pages} {064428} (\bibinfo {year}
  {2011})}\BibitemShut {NoStop}%
\bibitem [{\citenamefont {He}\ \emph {et~al.}(2012)\citenamefont {He},
  \citenamefont {Wang},\ and\ \citenamefont {Chen}}]{He2011}%
  \BibitemOpen
  \bibfield  {author} {\bibinfo {author} {\bibfnamefont {J.~B.}\ \bibnamefont
  {He}}, \bibinfo {author} {\bibfnamefont {D.~M.}\ \bibnamefont {Wang}}, \ and\
  \bibinfo {author} {\bibfnamefont {G.~F.}\ \bibnamefont {Chen}},\ }\href
  {\doibase 10.1063/1.3694760} {\bibfield  {journal} {\bibinfo  {journal}
  {Appl. Phys. Lett.}\ }\textbf {\bibinfo {volume} {100}},\ \bibinfo {pages}
  {112405} (\bibinfo {year} {2012})}\BibitemShut {NoStop}%
\bibitem [{\citenamefont {Chapon}\ \emph {et~al.}(2011)\citenamefont {Chapon},
  \citenamefont {Manuel}, \citenamefont {Radaelli}, \citenamefont {Benson},
  \citenamefont {Perrott}, \citenamefont {Ansell}, \citenamefont {Rhodes},
  \citenamefont {Raspino}, \citenamefont {Duxbury}, \citenamefont {Spill},\
  and\ \citenamefont {Norris}}]{Chapon2011}%
  \BibitemOpen
  \bibfield  {author} {\bibinfo {author} {\bibfnamefont {L.}~\bibnamefont
  {Chapon}}, \bibinfo {author} {\bibfnamefont {P.}~\bibnamefont {Manuel}},
  \bibinfo {author} {\bibfnamefont {P.}~\bibnamefont {Radaelli}}, \bibinfo
  {author} {\bibfnamefont {C.}~\bibnamefont {Benson}}, \bibinfo {author}
  {\bibfnamefont {L.}~\bibnamefont {Perrott}}, \bibinfo {author} {\bibfnamefont
  {S.}~\bibnamefont {Ansell}}, \bibinfo {author} {\bibfnamefont
  {N.}~\bibnamefont {Rhodes}}, \bibinfo {author} {\bibfnamefont
  {D.}~\bibnamefont {Raspino}}, \bibinfo {author} {\bibfnamefont
  {D.}~\bibnamefont {Duxbury}}, \bibinfo {author} {\bibfnamefont
  {E.}~\bibnamefont {Spill}}, \ and\ \bibinfo {author} {\bibfnamefont
  {J.}~\bibnamefont {Norris}},\ }\href@noop {} {\bibfield  {journal} {\bibinfo
  {journal} {Neutron News}\ }\textbf {\bibinfo {volume} {22}},\ \bibinfo
  {pages} {22} (\bibinfo {year} {2011})}\BibitemShut {NoStop}%
\bibitem [{\citenamefont {Strocov}\ \emph {et~al.}(2010)\citenamefont
  {Strocov}, \citenamefont {Schmitt}, \citenamefont {Flechsig}, \citenamefont
  {Schmidt}, \citenamefont {Imhof}, \citenamefont {Chen}, \citenamefont
  {Raabe}, \citenamefont {Betemps}, \citenamefont {Zimoch}, \citenamefont
  {Krempasky}, \citenamefont {Wang}, \citenamefont {Grioni}, \citenamefont
  {Piazzalunga},\ and\ \citenamefont {Patthey}}]{Strocov2010}%
  \BibitemOpen
  \bibfield  {author} {\bibinfo {author} {\bibfnamefont {V.~N.}\ \bibnamefont
  {Strocov}}, \bibinfo {author} {\bibfnamefont {T.}~\bibnamefont {Schmitt}},
  \bibinfo {author} {\bibfnamefont {U.}~\bibnamefont {Flechsig}}, \bibinfo
  {author} {\bibfnamefont {T.}~\bibnamefont {Schmidt}}, \bibinfo {author}
  {\bibfnamefont {A.}~\bibnamefont {Imhof}}, \bibinfo {author} {\bibfnamefont
  {Q.}~\bibnamefont {Chen}}, \bibinfo {author} {\bibfnamefont {J.}~\bibnamefont
  {Raabe}}, \bibinfo {author} {\bibfnamefont {R.}~\bibnamefont {Betemps}},
  \bibinfo {author} {\bibfnamefont {D.}~\bibnamefont {Zimoch}}, \bibinfo
  {author} {\bibfnamefont {J.}~\bibnamefont {Krempasky}}, \bibinfo {author}
  {\bibfnamefont {X.}~\bibnamefont {Wang}}, \bibinfo {author} {\bibfnamefont
  {M.}~\bibnamefont {Grioni}}, \bibinfo {author} {\bibfnamefont
  {A.}~\bibnamefont {Piazzalunga}}, \ and\ \bibinfo {author} {\bibfnamefont
  {L.}~\bibnamefont {Patthey}},\ }\href {\doibase 10.1107/S0909049510019862}
  {\bibfield  {journal} {\bibinfo  {journal} {J. Sync. Rad.}\ }\textbf
  {\bibinfo {volume} {17}},\ \bibinfo {pages} {631} (\bibinfo {year}
  {2010})}\BibitemShut {NoStop}%
\bibitem [{\citenamefont {Miller}\ and\ \citenamefont
  {Love}(1967)}]{Miller1967}%
  \BibitemOpen
  \bibfield  {author} {\bibinfo {author} {\bibfnamefont {S.~C.}\ \bibnamefont
  {Miller}}\ and\ \bibinfo {author} {\bibfnamefont {W.~F.}\ \bibnamefont
  {Love}},\ }\href@noop {} {\emph {\bibinfo {title} {Tables of Irreducible
  Representations of Space Groups and Co-Representations of Magnetic Space
  Groups}}}\ (\bibinfo  {publisher} {Pruett Press},\ \bibinfo {year}
  {1967})\BibitemShut {NoStop}%
\bibitem [{\citenamefont {Schoop}\ \emph {et~al.}(2018)\citenamefont {Schoop},
  \citenamefont {Topp}, \citenamefont {Lippmann}, \citenamefont {Orlandi},
  \citenamefont {M{\"u}chler}, \citenamefont {Vergniory}, \citenamefont {Sun},
  \citenamefont {Rost}, \citenamefont {Duppel}, \citenamefont {Krivenkov},
  \citenamefont {Sheoran}, \citenamefont {Manuel}, \citenamefont {Varykhalov},
  \citenamefont {Yan}, \citenamefont {Kremer}, \citenamefont {Ast},\ and\
  \citenamefont {Lotsch}}]{Schoopeaar2317}%
  \BibitemOpen
  \bibfield  {author} {\bibinfo {author} {\bibfnamefont {L.~M.}\ \bibnamefont
  {Schoop}}, \bibinfo {author} {\bibfnamefont {A.}~\bibnamefont {Topp}},
  \bibinfo {author} {\bibfnamefont {J.}~\bibnamefont {Lippmann}}, \bibinfo
  {author} {\bibfnamefont {F.}~\bibnamefont {Orlandi}}, \bibinfo {author}
  {\bibfnamefont {L.}~\bibnamefont {M{\"u}chler}}, \bibinfo {author}
  {\bibfnamefont {M.~G.}\ \bibnamefont {Vergniory}}, \bibinfo {author}
  {\bibfnamefont {Y.}~\bibnamefont {Sun}}, \bibinfo {author} {\bibfnamefont
  {A.~W.}\ \bibnamefont {Rost}}, \bibinfo {author} {\bibfnamefont
  {V.}~\bibnamefont {Duppel}}, \bibinfo {author} {\bibfnamefont
  {M.}~\bibnamefont {Krivenkov}}, \bibinfo {author} {\bibfnamefont
  {S.}~\bibnamefont {Sheoran}}, \bibinfo {author} {\bibfnamefont
  {P.}~\bibnamefont {Manuel}}, \bibinfo {author} {\bibfnamefont
  {A.}~\bibnamefont {Varykhalov}}, \bibinfo {author} {\bibfnamefont
  {B.}~\bibnamefont {Yan}}, \bibinfo {author} {\bibfnamefont {R.~K.}\
  \bibnamefont {Kremer}}, \bibinfo {author} {\bibfnamefont {C.~R.}\
  \bibnamefont {Ast}}, \ and\ \bibinfo {author} {\bibfnamefont {B.~V.}\
  \bibnamefont {Lotsch}},\ }\href {\doibase 10.1126/sciadv.aar2317} {\bibfield
  {journal} {\bibinfo  {journal} {Sci. Adv.}\ }\textbf {\bibinfo {volume} {4}}
  (\bibinfo {year} {2018}),\ 10.1126/sciadv.aar2317}\BibitemShut {NoStop}%
\bibitem [{\citenamefont {Vergniory}\ \emph {et~al.}(2018)\citenamefont
  {Vergniory}, \citenamefont {Elcoro}, \citenamefont {Orlandi}, \citenamefont
  {Balke}, \citenamefont {Chan}, \citenamefont {Nuss}, \citenamefont
  {Schnyder},\ and\ \citenamefont {Schoop}}]{Vergniory2018}%
  \BibitemOpen
  \bibfield  {author} {\bibinfo {author} {\bibfnamefont {M.~G.}\ \bibnamefont
  {Vergniory}}, \bibinfo {author} {\bibfnamefont {L.}~\bibnamefont {Elcoro}},
  \bibinfo {author} {\bibfnamefont {F.}~\bibnamefont {Orlandi}}, \bibinfo
  {author} {\bibfnamefont {B.}~\bibnamefont {Balke}}, \bibinfo {author}
  {\bibfnamefont {Y.-H.}\ \bibnamefont {Chan}}, \bibinfo {author}
  {\bibfnamefont {J.}~\bibnamefont {Nuss}}, \bibinfo {author} {\bibfnamefont
  {A.~P.}\ \bibnamefont {Schnyder}}, \ and\ \bibinfo {author} {\bibfnamefont
  {L.~M.}\ \bibnamefont {Schoop}},\ }\href@noop {} {\bibfield  {journal}
  {\bibinfo  {journal} {The Euro. Phys. J. B}\ }\textbf {\bibinfo {volume}
  {91}},\ \bibinfo {pages} {213} (\bibinfo {year} {2018})}\BibitemShut
  {NoStop}%
\end{thebibliography}%


%
\end{document}